\documentclass[fleqn,usenatbib]{mnras}

\usepackage{newtxtext,newtxmath}

\usepackage[T1]{fontenc}

\DeclareRobustCommand{\VAN}[3]{#2}
\let\VANthebibliography\thebibliography
\def\thebibliography{\DeclareRobustCommand{\VAN}[3]{##3}\VANthebibliography}

\usepackage{graphicx}	
\usepackage{amsmath}	
\usepackage{subcaption} 
\usepackage{xcolor}
\usepackage{multirow}
\usepackage{pdflscape}

\title[MWA Phase II Delay Spectrum Limits]{New EoR Power Spectrum Limits From MWA Phase II Using the Delay Spectrum Method and Novel Systematic Rejection}

\author[M. Kolopanis et al.]{Matthew Kolopanis$^{1}$\thanks{E-mail: Matthew.Kolopanis@asu.edu}, Jonathan C. Pober$^{2}$, Daniel C. Jacobs$^{1}$, Samantha McGraw$^{2}$
\\
\\
$^{1}$School of Earth and Space Exploration, Arizona State University, Tempe, AZ, USA\\
$^{2}$Department of Physics, Brown University, Providence, RI, USA\\
}

\date{Accepted XXX. Received YYY; in original form ZZZ}

\pubyear{2021}

\begin{document}
\label{firstpage}
\pagerange{\pageref{firstpage}--\pageref{lastpage}}
\maketitle

\begin{abstract}
We present an analysis of Epoch of Reionization data from Phase II of the Murchison Widefield Array using the \texttt{simpleDS} delay spectrum pipeline. Prior work analyzed the same observations using the FHD/$\varepsilon$ppsilon imaging pipeline, and so the present analysis represents the first time that both principal types of 21\,cm cosmology power spectrum estimation approaches have been applied to the same data set.  Our limits on the 21\,cm power spectrum amplitude span a range in $k$ space of $|k| < 1~h_{100}{\rm Mpc}^{-1}$ with a lowest measurement of $\Delta^2(k) \leq$ $4.58\times10^3$\,mK$^2$ at $k = 0.190\,h_{100}\rm{Mpc}^{-1}$ and $z = 7.14$.  In order to achieve these limits, we need to mitigate a previously unidentified common mode systematic in the data set.  If not accounted for, this systematic introduces an overall \emph{negative} bias that can make foreground contaminated measurements appear as stringent, noise-limited constraints on the 21\,cm signal amplitude. The identification of this systematic highlights the risk in modeling systematics as positive-definite contributions to the power spectrum and in ``conservatively'' interpreting all measurements as upper limits.  

\end{abstract}




\section{Introduction}

Detecting the highly-redshifted 21\,cm signal from neutral hydrogen is a major goal of modern cosmology.  Multiple experiments targeting such a detection are now either complete or underway, including 
The Donald C. Backer Precision Array for Probing the Epoch of Reionization \citep[PAPER;][]{parsons_et_al_2010,kolopanis_et_al_2019}, 
the Giant Metrewave Radio Telescope \citep[GMRT;][]{swarup_et_al_1991,paciga_et_al_2013},
the Low-Frequency Array \citep[LOFAR;][]{van_haarlem_et_al_2013,mertens_et_al_2020},
the Hydrogen Epoch of Reionization Array
\citep[HERA;][]{deboer_et_al_2017,hera_collaboration_2022},
and the Murchison Widefield Array \citep[MWA;][]{tingay_et_al_2013a,wayth_et_al_2018,trott_et_al_2020},
all of which are radio interferometers with the aim of measuring the power spectrum of spatial fluctuations in the hydrogen signal during the cosmological Epoch of Reionization.\footnote{Many other approaches for pursuing this signal exist, including measurements of the sky-averaged ``global" signal \citep[e.g.][]{bowman_et_al_2018} and measurements aimed at other redshifts \citep[e.g.][]{eastwood_et_al_2019}.}  The 21\,cm signal must be separated from extremely bright foreground emission based on the distinct spectral behavior of the two emission mechanisms (spectrally structured for the 21\,cm signal and smooth for the foregrounds).  Teams from each experiment have developed a number of techniques to enable the extraction of the 21\,cm signal, including novel approaches to calibration, systematic mitigation, and power spectrum estimation.  

\citet{morales_et_al_2019} show that, despite the intricacies of each approach, techniques used for power spectrum (PS) estimation can largely be classified into two categories: ``reconstructed sky'' estimators, which attempt to correct for the response of the radio interferometer to recover the power spectrum of the ``true'' sky, and ``measured sky'' estimators, which make no correction for the telescope response and instead recover the power spectrum of the sky ``as seen'' by the telescope.  Through simulations, \citet{morales_et_al_2019} demonstrate the signature of foreground contamination in the measured power spectra are distinct for these two classes of estimators.  Moreover, they describe how errors in the calibration of the telescope affect each class of estimator in distinct ways.  To-date, however, no one has applied both kinds of estimators to the exact same data set to study their real-world differences.

In this paper, we present a measured sky analysis of data from Phase II of the MWA using the \texttt{simpleDS} pipeline originally described by \citet{kolopanis_et_al_2019} and applying it to a dataset calibrated by \citet{li_et_al_2019}.  This analysis is complementary to that of \citet{li_et_al_2019}, which used a reconstructed sky-estimator (the FHD/$\varepsilon$ppsilon pipeline; \citealt{barry_et_al_2019a}).  In practice, however, our analyses focus on different subsets of the same data set.  In the present analysis, we focus on the three shortest spacings (14\,m, 24.24\, and 28\,m baseline lengths), as they have both the greatest intrinsic sensitivity to the cosmological signal and the largest amount of redundancy. In the \citet{li_et_al_2019} analysis of this set, for reasons we describe below, the 14\,m baselines were excluded from the final power spectrum estimation, while a large number of longer baselines (up to 80\,m) were also included.


In this latest analysis, we find that many of our baselines (particularly, but not only, those with an east-west orientation) have a strong common mode signal. This signal has not been previously reported and, if not mitigated, adds a statistically significant offset to power spectrum measurements.  This contamination is most visible in noise-dominated delay modes, where it shows up as a time-stable phase offset.  We also find that the angle of the phase offset varies from baseline to baseline. As described in \S\ref{sec:dspec}, our power spectrum estimator cross-multiplies nominally redundant baselines; because of the differing phase angles of the common mode signal, the result is a \emph{negative power spectrum value}, which in certain cases can be quite large.  This discovery necessitated development of a new set of metrics identifying this common mode signal and flagging contaminated baselines (\S\ref{sec:quality2}).  Removing contaminated data not only eliminates the negative power spectrum points, but (in some cases) we also find modes where a modest negative bias had previously obscured faint foreground contamination, making what should have been a high-significance detection of a non-EoR signal appear to be a noise-limited measurement.  This result illustrates the subtlety of systematic errors in 21\,cm PS analyses --- contaminating signals can lead to \emph{negative} biases, not just positive offsets --- and emphasizes the importance of data quality control prior to PS estimation.

The structure of this paper is as follows.
In \S\ref{sec:dspec}, we present an overview of the delay spectrum approach to PS estimation, highlighting the key points needed to understand the appearance of negative PS signals.
In \S\ref{sec:obs}, we describe the MWA observations used in our analysis, and in \S\ref{sec:data}, we present our pre-processing steps (including an overview of the steps applied in \citealt{li_et_al_2019} and a more detailed discussion of the steps new to this analysis).  In \S\ref{sec:sims}, we present an independent simulation of our data that is used to validate our analysis, and in \S\ref{sec:simpleDS}, we present the \texttt{simpleDS} pipeline in more detail.  We present our results (including our upper limits on the 21\,cm PS) in \S\ref{sec:results} and conclude with a discussion in \S\ref{sec:discussion}. Throughout the work, we assume the Planck 2015 cosmology \citep{planck_2015_XIII}.

\section{Delay Spectra}
\label{sec:dspec}

The common mode systematic and the potential for negative power spectrum points are a key feature of the analysis presented here.  As such, it is important to review the delay spectrum approach in general to see how such effects can come about.

The delay spectrum power spectrum method relies on the observation that a single baseline, to a close approximation, can sample a spectrum of a single spatial wavemode.  For 21\,cm cosmology, then, a Fourier transform of a single baseline's visibilities along the frequency dimension yields a series of line-of-sight wavemodes measured at a single transverse (i.e. spatial) mode.  This operation is subject to a number of constraints and approximations which we describe below.
  See \citet{parsons_and_backer_2009} and \citet{parsons_et_al_2012a,parsons_et_al_2012b} for more information, as well as \citet{liu_and_shaw_2020} for a pedagogical review.

The delay transform is calculated as the Fourier transform of the interferometric visibilities $V$ from a \emph{single baseline}, $b$, across the frequency axis:
\begin{equation}
\label{eq:dtransform}
\tilde{V}_b(\tau) = \int \mathrm{d}\nu~V_b(\nu)~e^{2\pi i\nu\tau},
\end{equation}
where $\tau$ is the delay, i.e., the time between the arrival of the signal from one antenna of the baseline relative to the other.  The principal source of such a delay is the \emph{geometric delay}:
\begin{equation}
\label{eq:taudef}
\tau_g = \frac{\mathbf{b}\cdot\hat{s}}{c},
\end{equation}
where $\textbf{b}$ is the positional vector between the two antennas, $\hat{s}$ is the source vector, and $c$ is the speed of light.  Because radiation from different positions on the sky, $\hat{s}$, arrives with different geometric delays, measuring the delay transform of a baseline's visibilities can effectively provide a coarse, one-dimensional\footnote{The ``image" is 1D because only the source position in the direction parallel to the baseline vector will change the geometric delay; shifts in positions perpendicular to the baseline vector leave the delay unchanged.} image of the sky.  Celestial emission is confined to the largest realizable geometric delay between two antennas. This is colloquially known as the horizon, because only sources at the horizon can have the maximum delay.  Note that the relationship between the delay transform and geometric delay only holds for spectrally \emph{flat} emission; emission with inherent spectral structure (like the EoR signal) will exhibit power in high delay modes even if it comes from a direction with small geometric delay.  Delay modes beyond the horizon limit, where we do not expect to find smooth-spectrum foreground emission, are therefore said to be in the ``EoR window".  Delay modes smaller than the horizon value are in the ``foreground wedge'' (where the ``wedge'' shape arises when comparing baselines of different length, as longer baselines have longer maximum delays associated with the horizon).

\citet{parsons_et_al_2012a,parsons_et_al_2012b} show that, because the spectral structure of the 21\,cm signal probes the line-of-sight structure of the EoR, a set of delay transformed visibilities can be used as an estimator of the cosmological power spectrum:
\begin{equation}
\label{eq:pspec_approx}
|\tilde V_b(\tau)|^2 \propto P(k),
\end{equation}
where the proportionality factor is a cosmological normalization that depends on the center frequency (i.e. redshift) and bandwidth of the observation, as well as the primary beam of the antennas.  In this work, we follow this approach for estimating the cosmological power spectrum, using the code \texttt{simpleDS}\footnote{https://github.com/rasg-affiliates/simpleDS} \citep{kolopanis_et_al_2019}, which we describe further in \S\ref{sec:simpleDS}.  

Directly following Equation \ref{eq:pspec_approx} and squaring the delay-transformed visibilities, however, will leave a noise bias in the data (i.e. once squared, the zero-mean noise becomes positive-definite and will not integrate down with the accumulation of more data.  Several approaches exist to avoid this noise bias, including direct subtraction of an independent noise estimate (e.g. from Stokes V).  Our approach is to cross-multiply statistically independent measurements using, for example, adjacent time samples or repeated observations.  When antennas are arranged in a grid, such as with the MWA Phase II setup, this can be done by cross multiplying different pairs of antennas with the same physical vector separation. These are sometimes called \emph{redundant} baselines.

Since redundant baselines (nominally) measure the same sky signal but with different noise realizations, this approach will produce an estimate of the power spectrum where the noise is still zero-mean and thus will continue to integrate down as more data is accumulated.  Although unphysical for a real-valued sky signal, this approach can lead to negative power spectrum values, either because the noise is zero mean or because of the presence of a non-redundant systematic which appears out of phase on the baselines being cross-multiplied.  We find exactly such a systematic in this analysis, the identification and mitigation of which is discussed further in \S\ref{sec:quality2}.

\section{Observations}
\label{sec:obs}

\begin{figure}
    \centering
    \includegraphics[width=\columnwidth]{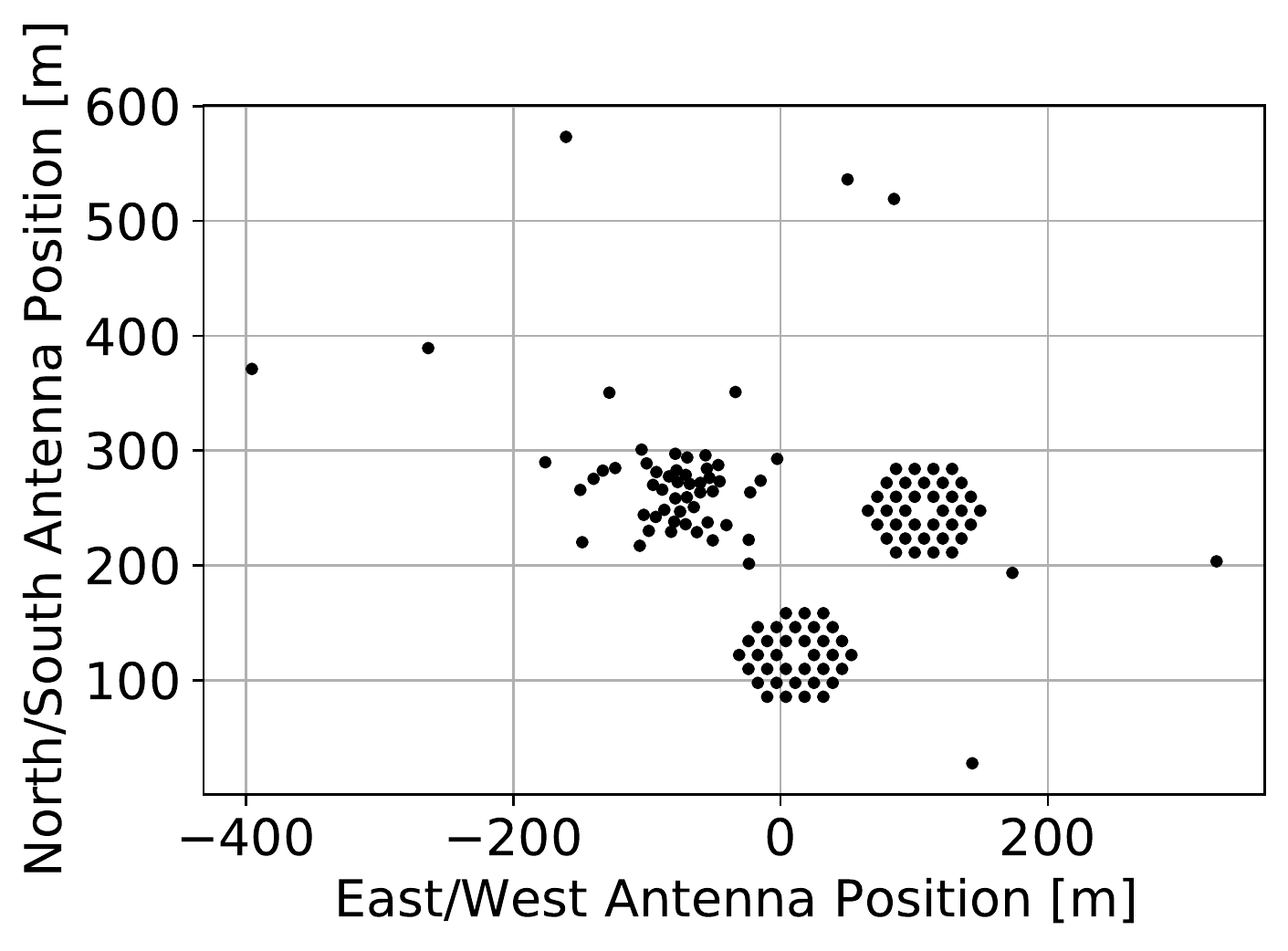}
    \caption{The Antenna positions of the Murchison Widefield Array (MWA) Phase II compact configuation in an East, North, Up (ENU) coordinate system relative the array location.}
    \label{fig:layout}
\end{figure}

Phase II of the MWA \citep{wayth_et_al_2018} consists of 256 ``tiles" of antennas, each having 16 dual-polarization dipoles arranged in a 4$\times$4 layout. The 16 tile signals are fed into an analog beamformer. The dual polarization outputs of the beamformer are digitized and correlated.  Limitations in the digitizer and correlator permit observations with only 128 tiles at a time, and as such the array operates in two modes: a compact array and an extended array, each of which uses a subset of the available tiles.  The compact array consists of two hexagonal cores and 56 pseudo-randomly distributed tiles, the layout of which is shown in Figure \ref{fig:layout}.  

\citet{li_et_al_2019} analyzed 40 hours of compact configuration observations recorded between October 15 and December 15, 2016. These recordings targeted a field referred to as ``EoR0" at RA 0\,h, Dec -27$^\circ$, in the instrument's high band (167 - 197 MHz, corresponding to redshifts $z$ = 7.5 - 6.2 in the 21 cm line).  Individual dipoles in the tile beamformer can be phased to form primary beams toward the desired sky direction. During observations the target region was allowed to drift through the field of view with a new pointing established roughly every 30 minutes to keep the field near the center of the primary beam.  Though the observation used eight pointings to track for roughly four hours each night (five of which we used in the final PS of \citealt{li_et_al_2019}), here we consider only the zenith scans, leaving the integration of steered pointings into a delay spectrum analysis for future work. At present \texttt{simpleDS} is not set up to consistently treat a primary beam that changes over the course of the observation.   This zenith subset amounts to approximately 6.2 hours of data, observed over the course of 12 nights. The set  is divided into 200 files (each file is referred to as an ``observation'') each 112 seconds long.  Data were recorded with a time resolution of 0.5 seconds and a frequency resolution of 40 kHz.

The MWA spectrum is formed in a two step process which leads to a distinctive passband and channel structure.  The MWA observes a simultaneous bandwidth of 30.72\, MHz selected from a broader range of 80-300\,MHz.  Signals are first divided into 24 coarse channels, each 1.28\,MHz wide.  Each coarse channel is further divided into 32 fine frequency channels in a second stage Fourier transform, each with a resolution of 40 kHz.  However, aliasing between coarse channels, where signals wrap within coarse bands, means that edges of each coarse channel must be flagged \citep{prabu_et_al_2015}.  The data therefore have a periodic (every 1.28\,MHz) gap, which leads to prominent artifacts in the Fourier transforms used in our later analyses (referred to subsequently as the ``coarse band harmonics''). These are treated with a combination of modeling and avoidance as described below.

\section{Data Processing}
\label{sec:data}
The data analyzed here are the calibrated and flagged outputs resulting from the analysis by \citet{li_et_al_2019}; the reader is referred there for a detailed description. What follows is a compressed summary of that procedure.  In summary the process consisted of three steps: pre-processing with \texttt{COTTER} (\S\ref{sec:preprocessing}); calibration with FHD and \texttt{OMNICAL} (\S\ref{sec:calibration}); and a first stage of data quality cuts (\S\ref{sec:quality}).  These steps result in a selection of data with interference flagged and calibrations set to a repeatable phase and gain scale across many nights and baselines. On top of these steps we add: baseline selection (\S\ref{sec:baselines}), Local Sideral Time (LST) averaging across multiple nights (\S\ref{sec:lst}), and a second stage of data quality cuts (\S\ref{sec:quality2}).

\subsection{Pre-Processing}
\label{sec:preprocessing}
Pre-processing is performed by the \texttt{COTTER} algorithm, which has three main functions: 
(1) ingesting the custom MWA correlator file format and returning a \texttt{uvfits} file; 
(2) performing an initial round of radio-frequency interference (RFI) flagging using the \texttt{AOFLAGGER} algorithm \citep{offringa_et_al_2010,offringa_et_al_2012}; and
(3) applying time-averaging from 0.5 seconds to 2.0 seconds to reduce data volume.

\subsection{Calibration}
\label{sec:calibration}

Antenna-based gain calibration corrects for amplitude and phase differences across the array using a known sky model. The process, described in \citet{li_et_al_2019}, begins with a per-frequency calibration using FHD \citep{sullivan_et_al_2012,barry_et_al_2019a}, which uses simulated visibilities created from a sky and instrument model to solve for the gains that minimize least squares difference with data.  In \citet{li_et_al_2019} and other FHD analyses of EoR data, only baselines longer than $\sim$80\,m are used in this calibration process.  Because the sky model does not include any diffuse component, short baselines (where diffuse emission contributes a significant amount of the overall flux density measured) do not calibrate well.  Preliminary results from \citet{byrne_thesis} suggest that including a diffuse sky map enables these baselines to be used in calibration, but further research is necessary.  Here we use the calibration solutions from \citet{li_et_al_2019} which were produced using the 80\,m cut.

After the initial per-frequency calibration, the gains for the tiles within the two hexagonal cores of the array are then further updated using the redundancy-based \texttt{lincal} algorithm \citep{liu_et_al_2010} implemented by the \texttt{OMNICAL} package \citep{zheng_et_al_2014}.  Lastly, bandpasses for all antennas in the array are calibrated using antenna auto-correlations as described in \citet{li_et_al_2019}. This step mitigates known artifacts in the per-frequency gains and also fits reflections in the signal chain.  

\subsection{Quality Cuts: First Stage}
\label{sec:quality}

Though the Murchison Radio Observatory is one of the most isolated and protected radio environments in the world, transmitters still lead to low level interference. Long distance propagation mechanisms such as reflections from airplanes or satellites or tropospheric ducting mean that interfering signals are strongly variable, sometimes appearing strongly for brief moments. Several methods have been devised to detect and flag these types of signals. Flagging itself can introduce strong correlation between delay spectrum modes and additional flagging is sometimes necessary to remove observations with excessively high delay spectrum residuals.

The interference flagging protocol used by  \citet{li_et_al_2019}, used three quality metrics. The Sky-Subtracted Incoherent Noise Spectrum (SSINS) metric uses cross correlations to detect, faint, broad-band, interference like digital TV. This method has been shown to offer an improvement over the initial round of flagging done by \texttt{cotter} which uses \texttt{AOFlagger}, at least when run with the default settings \citep{wilensky_et_al_2019}.  Flags based on SSINS were selected by \citet{li_et_al_2019} at known frequencies, specific times, or entire observations based on inspection of metric distributions and prior knowledge of emitter bands. 

Some RFI was only detected by some of the antennas in the array such that it evades detection by SSINS which averages across all baselines. A sensitive measure of variation across antennas is the $\chi^2$ produced by \texttt{OMNICAL} during the redundant calibration process. The \texttt{OMNICAL} $\chi^2$ is a measure of non-redundancy across nominally redundant baselines; elevated values of $\chi^2$ that are localized to specific frequency bands are indicative of this class of RFI.  All observations showing such signatures in their $\chi^2$ values have been excluded.

Lastly, the power spectrum itself was used as a metric. Using FHD/$\varepsilon$ppsilon, a 2D cylindrical PS was calculated for each 112\,s observation.  The noise level in any individual observation is orders of magnitude above any predicted background model but still lower than the sensitivity of all the previous metrics. Observations showing power in the ``EoR window" well above the mean are flagged in their entirety. 

These quality cuts flagged a number of samples at specific times and frequencies and excluded 23 files completely. The remaining data set included 177 observations totalling approximately 5.5 hours of data.

\subsection{Baseline Selection}
\label{sec:baselines}

Unlike reconstructed sky estimators like FHD which combine data from many times and baselines to achieve the highest possible fidelity, measured sky estimators like \texttt{simpleDS} can operate on a per-baseline basis; each individual baseline can be used to obtain an independent estimate of the power spectrum; \citealt{parsons_et_al_2012a,parsons_et_al_2012b}). The advantageous simplicity of this approach has motivated several arrays to arrange antennas in a grid to maximize the number of times a few baseline vectors are measured. During the Phase II upgrade the MWA was outfitted with two hexagonally gridded 37 antenna subarrays (see Figure \ref{fig:layout}). In this study we only include correlations within these two hexes, a total of 72 antennas and 648 baselines.

\begin{figure}
    \centering
    \includegraphics[width=\columnwidth]{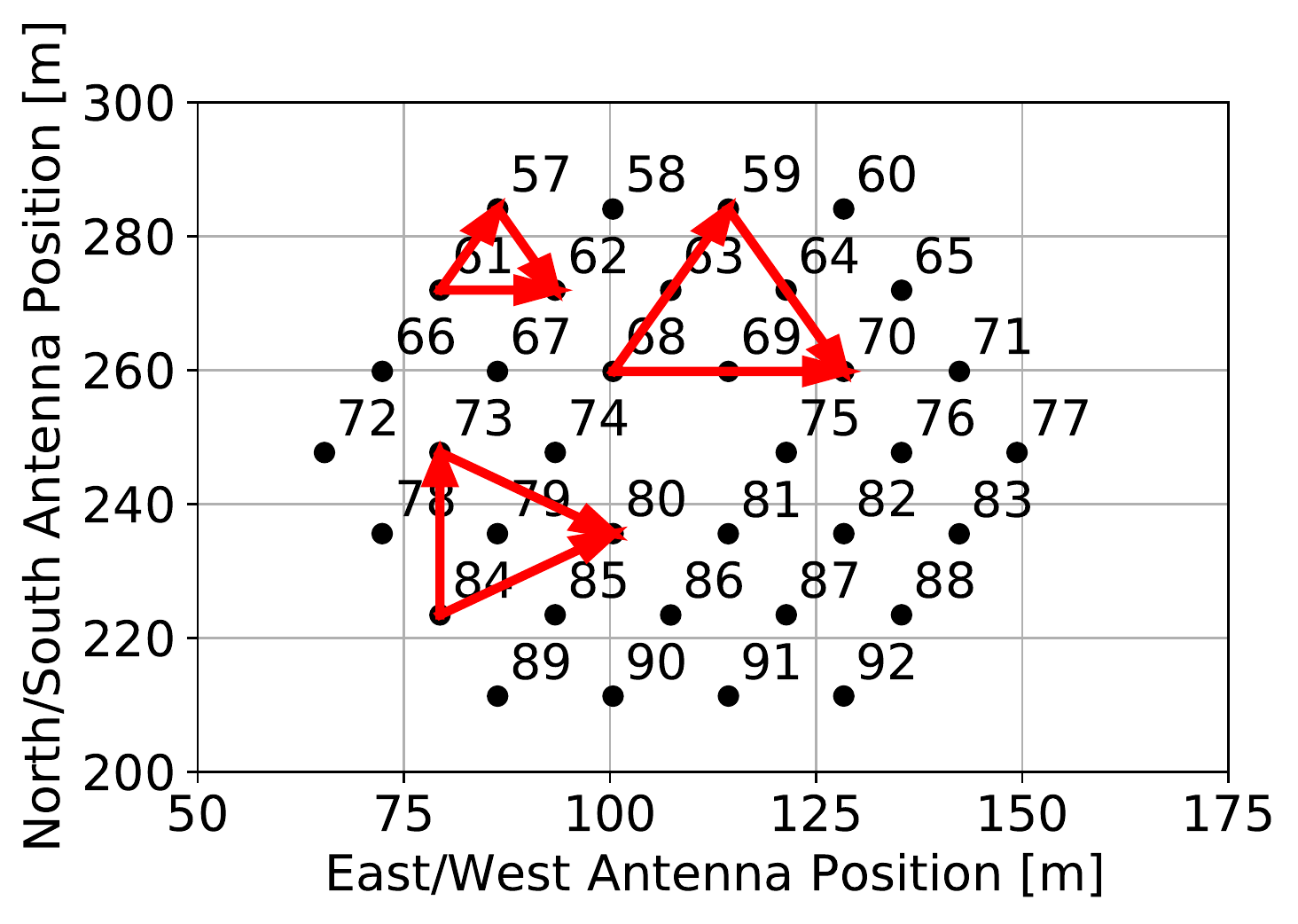}
    \caption{The 9 redundant baseline vectors used in this analysis plotted over one of the redundant array hexes from Figure \ref{fig:layout}. Lengths are 14\,m, 24.24\,m, and 28\,m,  with three orientations for each length.}
    \label{fig:bl_select}
\end{figure}

The delay power spectrum approach allows for an exploration of data sets at a much finer grained level; each baseline vector produces a power spectrum. But in the interests of fully inspecting the data set, further cuts were needed to focus the analysis.   Reasoning that since the 21cm background signal is predicted to be largest on the large spatial scales that are measured by the shortest baselines, we selected the 9 shortest baseline types in the redundant hexagonal cores of the MWA: 3 14\,m baselines (east-west, northeast, and southwest), 3 24.24 meter baselines, and 3 28 meter baselines, as illustrated in Figure \ref{fig:bl_select}.
Selecting only baselines of these 9 types leaves 418 individual baselines available for analysis.



This final data set has, in actuality, only partial overlap with the set analyzed in \citet{li_et_al_2019}.  As noted in \S\ref{sec:calibration}, FHD calibration uses only baselines longer than $\sim80$\,m given the lack of a diffuse sky model.  Furthermore, when averaging from 3D $k$ space down to a final 1D PS, \citet{li_et_al_2019} select only a range of $k_\perp$ values corresponding to baselines approximately 20 to 80\,m in length.  At the short baseline end, this cut is motivated by a significant increase in foreground power, both inside the wedge and the EoR window.  While the increase in power inside the wedge can be explained by the increasing amount of diffuse emission on the shortest baselines, the leakage into the EoR window requires a mechanism that throws power out of the wedge.  A definitive cause for this excess power has not yet been identified.  Due to these cuts in calibration and PS estimation, however, the shortest baselines at use in the present analysis have not been studied as well as other baselines in the array --- motivating, in part, the additional quality cuts described in \S\ref{sec:quality2}.


\subsection{LST Averaging}
\label{sec:lst}

\begin{figure*}
    \centering
    \includegraphics[width=\textwidth]{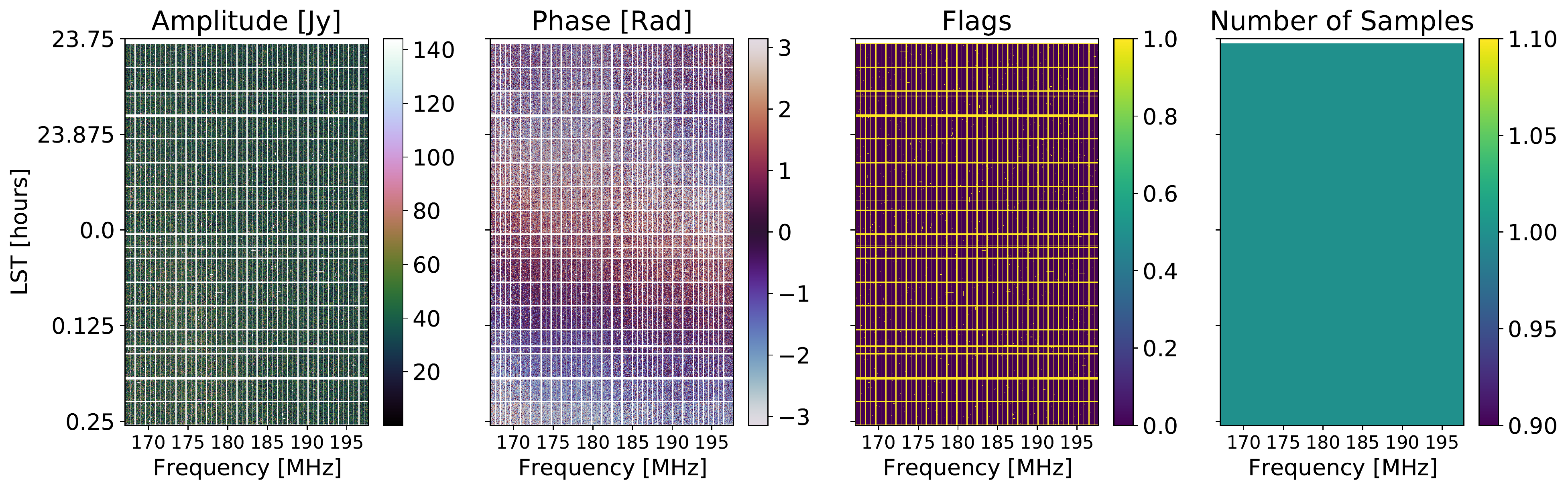}
    \includegraphics[width=\textwidth]{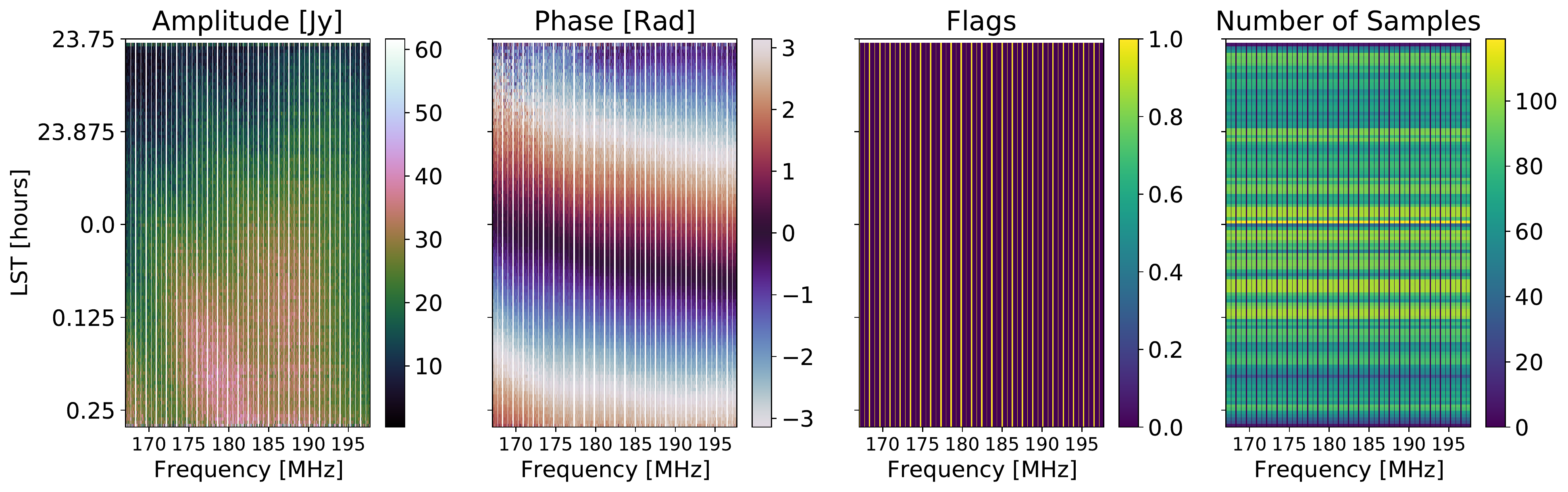}
    \caption{Top: (from left to right) the amplitude, phase, flag mask, and number of samples per bin for one night of data from one baseline.  The y-axis of each panel shows LST and the x-axis shows frequency.  Bottom: The same as above, but LST averaged across all nights of data.  Except for the flags from the coarse band structure, there are effectively no gaps in the data after LST averaging.
    }
    
    \label{fig:lstbin}
\end{figure*}

The data under study here are repeated observations of the RA$=0$ sky location recorded over the course of 12 nights. Averaging these observations decreases error due to noise, but just as importantly, averaging many nights helps fill in data missing due to interference flagging. The \texttt{simpleDS} algorithm calculates the delay transform with a simple Fast Fourier Transform along the frequency axis of each visibility; as such, it is important that the data be free from RFI flags, as flags in frequency will act as a point spread function in delay space and scatter foreground power to high delays.  Averaging multiple nights of data together, these gaps can be filled in.  

This gap-filling proceeds unevenly with more heavily flagged frequency channels being measured less often than others. This uneven sampling can still contribute to a less than ideal delay point spread function \citep{offringa_et_al_2019}. Inspecting the implied PSF given the channel sampling we conclude that these effects will be significantly lower than any residual seen in the present analysis. 

The LST binning procedure phases data to a common time grid and averages.  The grid of sidereal times is established spanning the LST range covered the data set (starting at LST=23h\,45m\,20.76s and ending at 00h\,16m\,21.84s for a total of just over 30 minutes) with a cadence of 16 seconds.  16 seconds is eight times longer than the data output by Cotter but much shorter than $\sim$10 minute fringe-crossing time for the longest baseline remaining in the data set (28\,m).  Each visibility in the data set is then matched to the closest LST bin and phased to the right ascension at the center of the bin using \texttt{pyuvdata} \citep{hazelton_et_al_2017}.  All visibilities falling within a bin are then averaged together.  

Figure \ref{fig:lstbin} shows the effects of the LST binning process.  The top row shows (from left to right) the amplitude, phase, flags, and number of samples for a single baseline for a single night of data with time on the $y$ axis and frequency on the $x$ axis.  (The number of samples panel is a single uniform color because, in the raw data, there is only one sample at each time and frequency).  The bottom row shows the same quantities after all 12 nights of data are combined in the LST average.  The flag mask now only shows the vertical streaks from the coarse band harmonics; nearly all the time dependent flags across the data set have been filled in.

Also of note is that, in the averaging process, we separate samples with even time indices from those with odd time indices (on the 2\,s cadence of the individual observations), and create two separate LST averaged sets.  These ``even'' and ``odd" sets serve as a jackknife of the data which can be used to empirically determine the noise level of the final data set.


\subsection{Quality Cuts: Second Stage}
\label{sec:quality2}
Following the above procedure we arrived at a data set with a small number of short baselines averaged over a dozen nights. As noted in \S\ref{sec:baselines}, these short baselines were not subjected to the same level of scrutiny as the longer baselines which form the bulk of the dataset used in \citet{li_et_al_2019}.  A new round of quality inspection revealed a new type of systematic and suggested another round of cuts. In this section, we will first describe the two defining features of this systematic: its stability in time (\S\ref{sec:systematic_time_stability}) and its variability between nominally redundant baselines (\S\ref{sec:systematic_bl_variance}).  We then present our approach for flagging it (\S\ref{sec:flagging_approach}) and the effect of this flagging strategy on the result (\S\ref{sec:flagging_results}).  In \S\ref{sec:systematic_discussion}, we briefly consider physical origins for this systematic signal, but ultimately do not identify the cause.

\subsubsection{Systematic Properties: Temporal Phase Stability}
\label{sec:systematic_time_stability}

\begin{figure*}
    \centering
    \includegraphics[width=\textwidth]{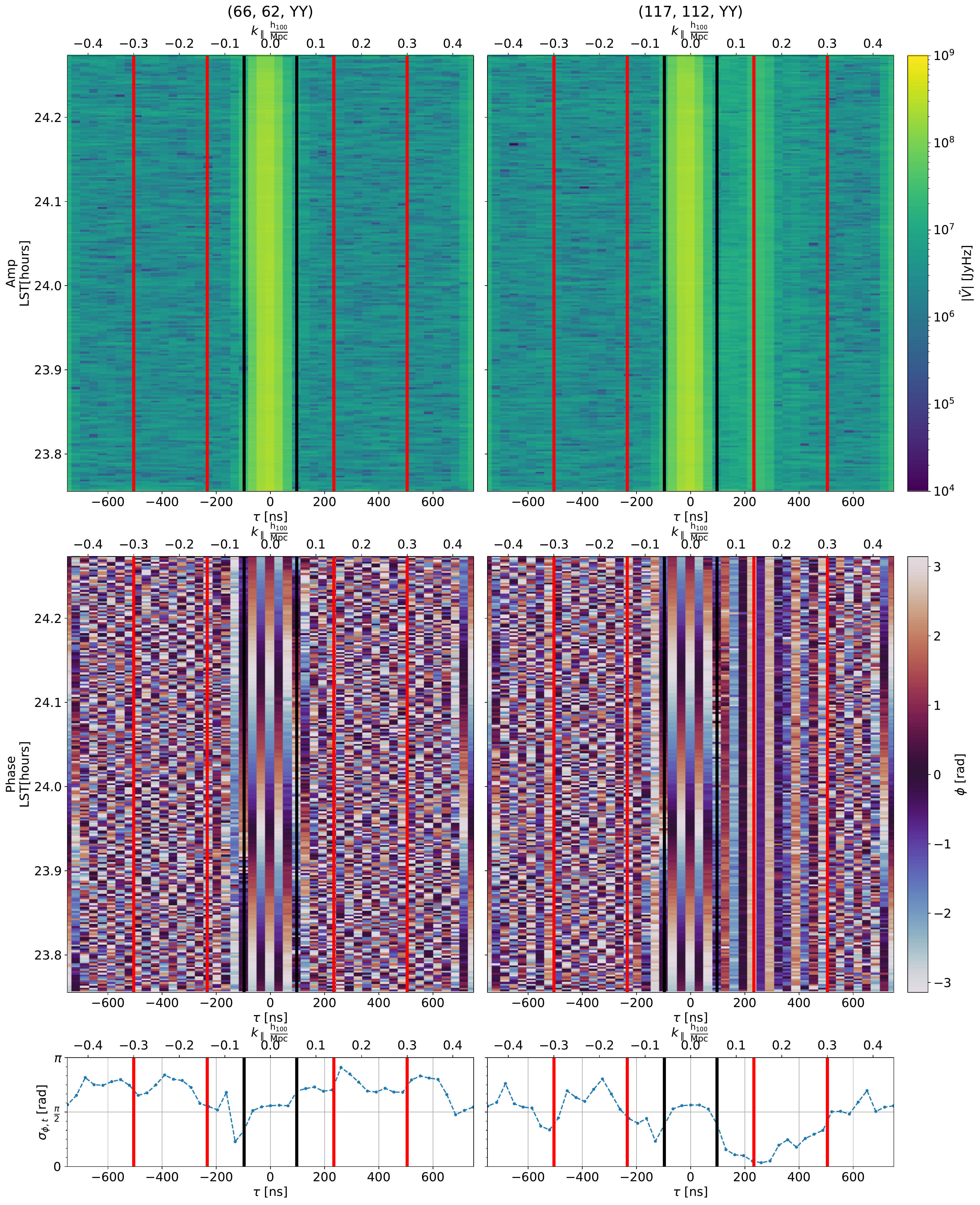}
    \caption{Comparison between a nominally uncontaminated baseline (66, 62), left column, and a highly contaminated baseline (117, 112), right column, in delay transformed amplitude (top), phase (middle), phase variance over time (bottom) over the LST range and delays used in this analysis. The region between the black bands indicate the geometric delay horizon for these baselines; additionally highlighted in red are the delay windows expected to be free from foreground leakage between coarse band ``ripples''. A phase-stable common mode systematic is present in the (117, 112) baseline at positive delays and even visible in the amplitude for this baseline.
    }
    \label{fig:systematic}
\end{figure*}

In order to understand the appearance of this newly identified common mode systematic, it is worthwhile to review the expected features of delay transformed visibilities.  The delay spectrum is a fully complex Fourier transform which yields both negative and positive delay values. As described in\S\ref{sec:dspec}, within the limits of the horizon, the position in delay space maps to the geometric delay of celestial emission arriving at the two antennas of the baseline. Positive delays indicate emission arriving from one direction, while negative originate from opposite side. As they move across the sky, a single source will follow a curve through delay space, originating at one horizon delay and setting on the other.\footnote{Whether positive delay maps to the eastern or western side is largely a sign/conjugation convention.}  However, given that our data set only spans 30 minutes of LST, little motion will be visible here. The large fringe pattern of the short baselines we analyze also leads to coarse delay resolution and only a small number of delay bins within the horizon.  

The left hand column of Figure \ref{fig:systematic} shows baseline (66,62), a ``normal" (i.e. systematic free) baseline, in delay space, where the $y$ axis corresponds to LST and the $x$ axis shows the value of delay in nanoseconds.  The top panel shows the amplitude of the delay transformed visibilities, while the middle panel shows the phase (the small panel in the bottom row of the plot will be discussed later).  As can be seen in the amplitudes, most of the power falls within the horizon as expected.  (The uptick of power right at the edges of the plot are harmonics of the low-delay foreground power, scattered by the flagged channels from the coarse bands.)  In \S\ref{sec:sims} we present the result of simulating visibilities from a foreground model. Among other things, these simulations suggest that foreground contamination will be weakest for modes of the delay spectrum between $\sim$225 and 500\,ns.  This range is indicated by the vertical red lines (one set for positive delays, and one set for negative). On this ``normal'' baseline, the measurements in this range appear dominated by noise.




Contrast this with baseline (117,112), on the right hand side of the plot, a prominent example of the suspect systematic. The signals inside the horizon on the two baselines appear consistent with each other, but here we see anomalous power at between 200 and 300\,ns, far outside the horizon lines. The systematic is particularly distinctive in the phase, where its stability stands in contrast to the uncorrelated noise seen in most baselines and delay modes of this type.  Note that the systematic phase shows less temporal evolution than the sky inside the horizon limits, where the phase changes as the Earth rotates.  The bottom row of Figure \ref{fig:systematic} illustrates the constancy of the phase with a plot of the average time variance of the phase, which we calculate using a definition adapted for circular boundary conditions \citep{mardia_1975}.  In our systematic contaminated baseline, the phase variance reaches very low values between $\sim$\,+200 and +300\,ns, in contrast to the baseline on the left, where the phase variance stays high throughout this range of delay space.

\subsubsection{Systematic Properties: Variance Across Redundant Baselines}
\label{sec:systematic_bl_variance}

The second defining feature of this systematic is what distinguishes it from sky signals (and, indeed, what makes it so problematic for a power spectrum analysis).  Whereas truly redundant baselines will measure the same sky signal, we observe our nominally redundant systematic contaminated baselines to have significantly different phase offsets.  The phase variance with time is \emph{low} while the variance from baseline to baseline is \emph{high}.  This situation is summarized in Figure \ref{fig:phasevar} which compares temporal phase variance ($y$ axis) against the variance of the time averaged phase across baselines ($x$ axis) for each delay bin. A noise simulation in grey is included for reference and delay modes are colored by  whether they are in the ranges expected for foregrounds (grey), cosmology (red for positive, blue for negative), or in-between (orangish-tan).  The dashed-lines and historgrams in the margins will be discussed in \S\ref{sec:flagging_approach}.

Our expectation is that delay modes dominated by sky will be repeatable. Depending on the baseline type in question, the amount of time included, and distribution of sources above the array, strong sky signals might have any range of time variance. But the time average phase, whatever it ends up being, will be consistent from baseline to baseline.  In delay modes dominated by noise we expect both temporal and baseline variation to be large.  Values with low time variance indicate stable signals while large variance baseline-to-baseline indicates signals which are not sky-like. Points in the bottom right quadrant of this plot are most like the systematic: stable in time, but unique to each pair of antennas.

\subsubsection{Flagging Approach}
\label{sec:flagging_approach}

Let us briefly discuss the impact of such a systematic on the power spectrum. This will help sharpen our mitigation strategy. First, consider a power spectrum formed by the conjugate product of delay spectra between two statistically independent baselines but nominally redundant baselines. Using baselines like (66,62), the ``normal'' example in Figure \ref{fig:systematic}, the delay channels dominated by noise will be zero mean while delay channels with strong sky power will be positive real. The time average of the product will always be consistent with a number greater than or equal to zero to within error bars. 

Next consider the product between two independent realizations of baselines like (117,112), which has the systematic.  If the phases of the systematic are very different between the two baselines, the resulting complex product between the two baselines will also be complex, with a real part that can be strongly negative.  The issue, of course, is that the sky is real-valued, and therefore cannot have a negative power spectrum.  As we have seen, the systematic we identify in our data indeed does vary significantly across nominally redundant baselines.  Flagging data contaminated by this systematic therefore is imperative for recovering physically meaningful limits on the 21\,cm PS.


\begin{figure*}
    \centering
    \includegraphics[width=0.7\textwidth]{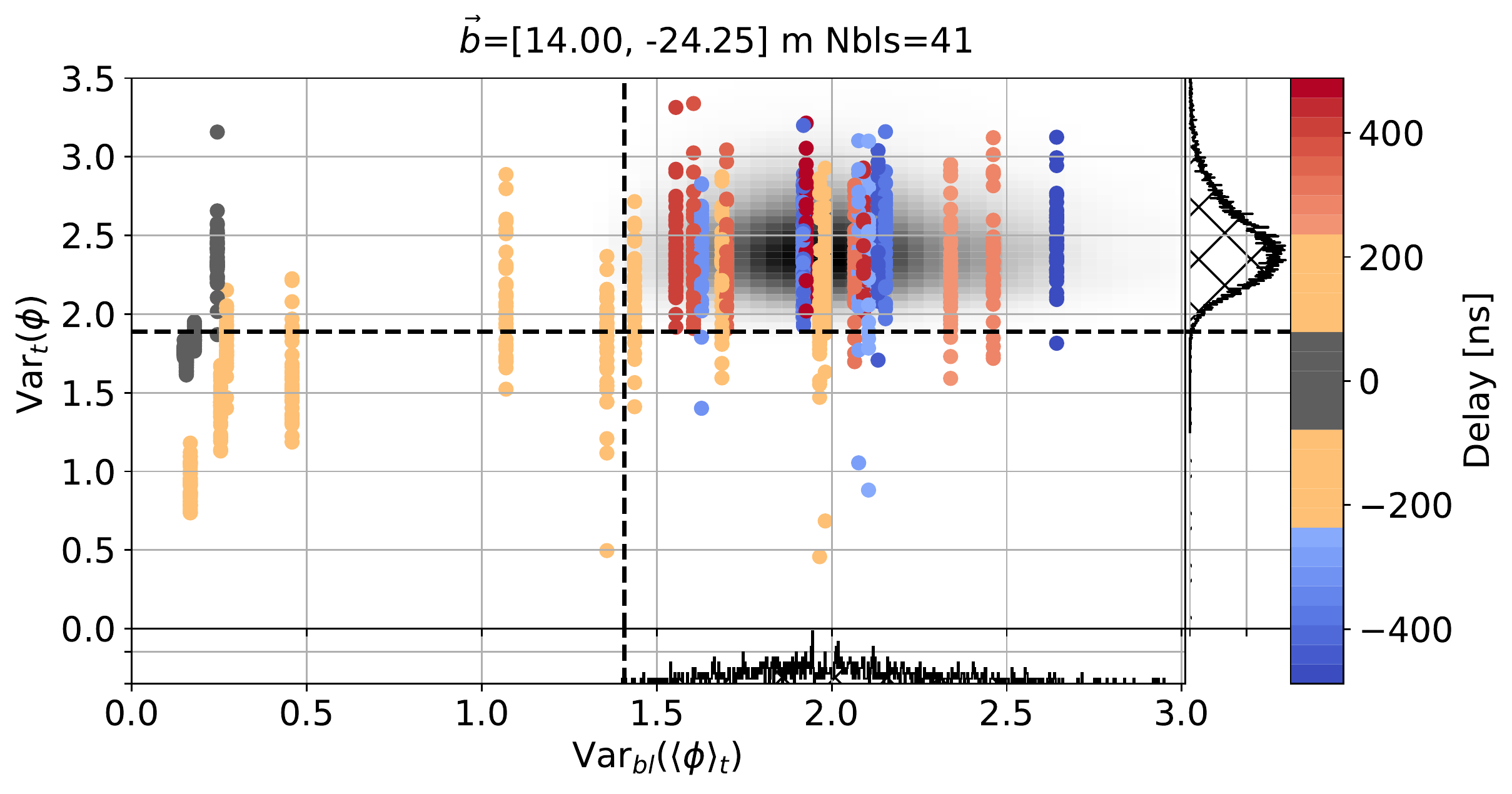}
    \includegraphics[width=0.7\textwidth]{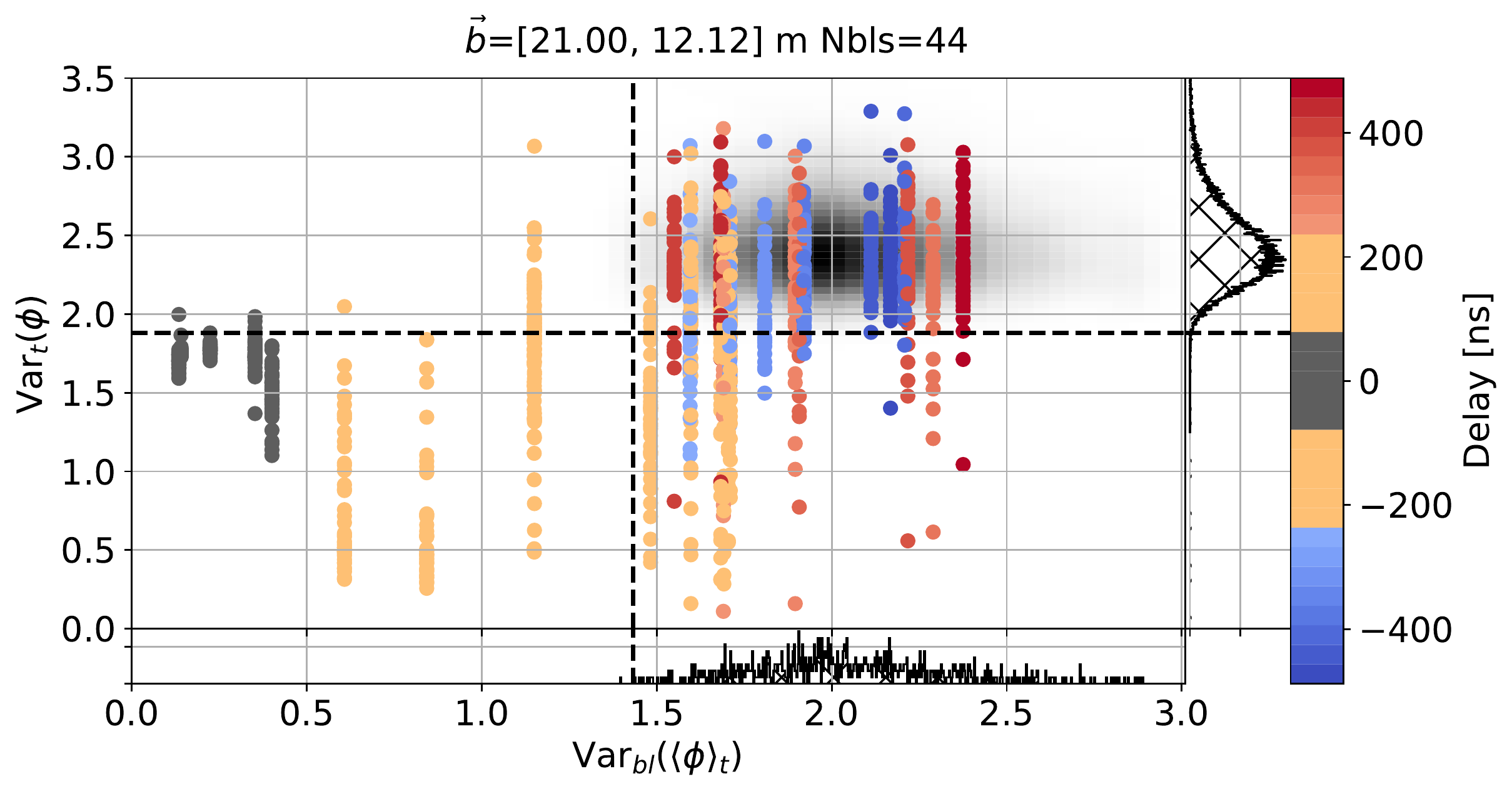}
    \includegraphics[width=0.7\textwidth]{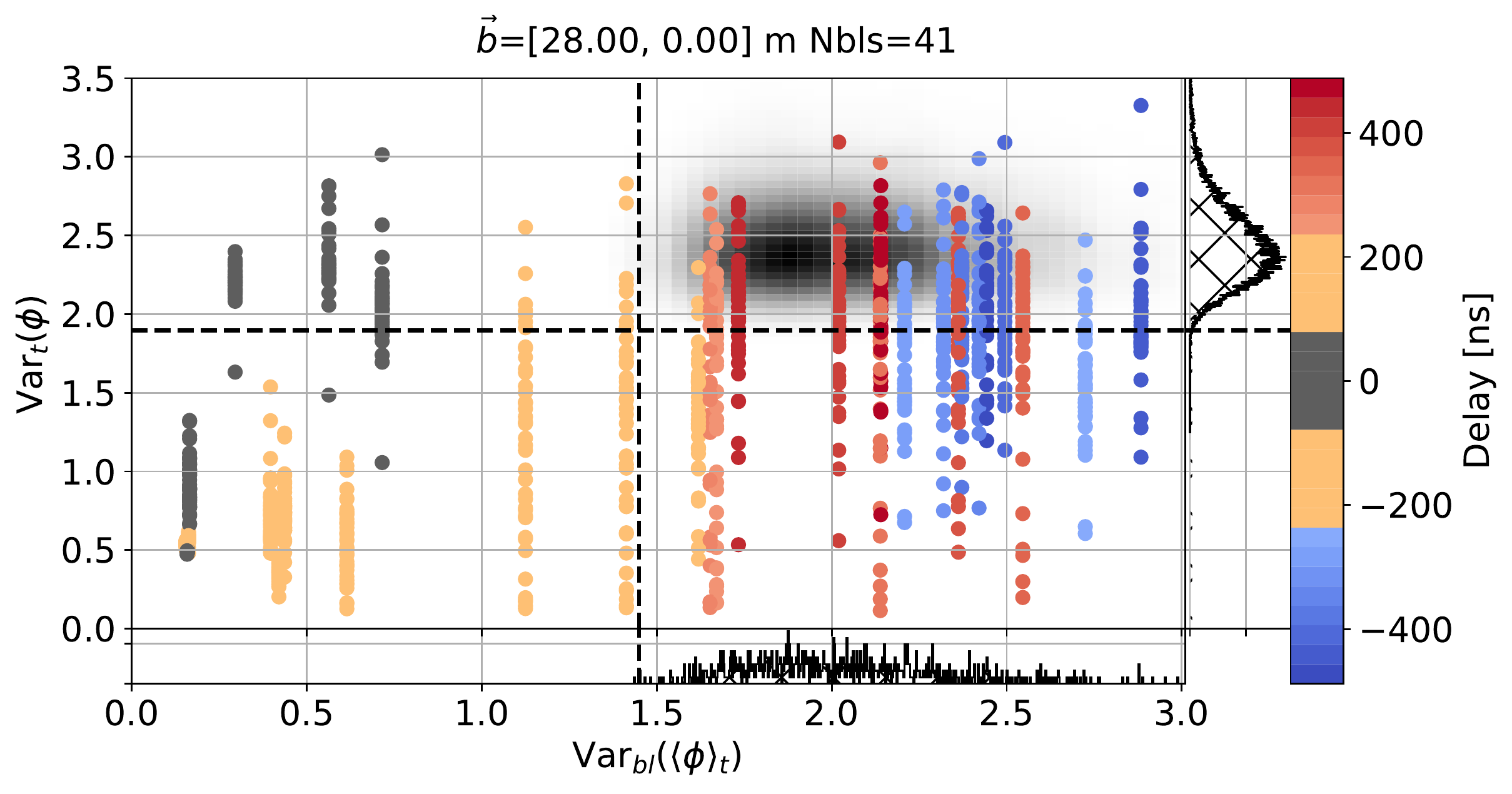}
    \caption{The space in which we isolate common mode contamination.  Each panel shows one baseline type for 'YY' polarization (the orientation is indicated in the panel title) and each point corresponds to the measurement in one delay bin from one baseline of the given type.  Color is used to indicate which delay bin the point comes from: gray points are inside the horizon limits, while red and blue points show delay modes from within the windows indicated with red lines in Figure \ref{fig:systematic} (red for positive delays, blue for negative).  Orangish-tan point are between the horizon and the windows of interest.  For each point, the $y$ location shows the variance of the phase versus time while the $x$ location shows the variance across all the baselines of the type under consideration.  (Hence the points appear clustered in vertical streaks, as each delay bin yields only one $x$ value per baseline \emph{type} while we get a different $y$ value for each baseline of that type.)  The gray-scale cloud is a kernel density estimation of the results of a noise simulation, i.e., it indicates where pure noise visibilities are found in this space.  The histograms on the right and bottom are 1D projections of the noise simulation distribution.  The vertical and horizontal dashed lines show the ``3$\sigma$'' effective cuts used to look for outliers (see text for details).  Common mode systematics show low time variance but high variance across nominally redundant baselines and thus contaminated modes are found in the lower right of the plot.
    }
    \label{fig:phasevar}
\end{figure*}

The defining signatures of the observed systematic can be summarized as a strong stable phase which exhibits baseline to baseline variance  (sometimes called ``non-redundancy'') beyond that expected for noise.  Formally, we define this using a noise simulation to calculate thresholds.  
In the simulation the visibility phases are replaced with uniformly distributed random numbers between $-\pi$ and $\pi$.  The results are indicated in the gray shaded regions in each panel of Figure \ref{fig:phasevar}, which are calculated using a kernel density estimator.  The 1D histograms on the right and bottom of each panel are projections of the simulated results onto the time-variance and baseline-variance axes, respectively. 
The $3\sigma$ effective bounds of the distributions from the noise simulation, i.e., the variance above which 99.86\% of the simulated points\footnote{Note that because we are only interested in finding tails on one side of our distribution, we use 99.86\% as opposed to 99.73\% as is generally associated with ``$3\sigma$''.} are marked by horizontal and vertical dashed lines across each panel of Figure \ref{fig:phasevar}.

The three baselines shown in Figure \ref{fig:phasevar} are sorted according to the number of points failing this check.  The top panel shows a relatively ``good'' baseline type, where only a few blue and red points fall into the lower right quadrant.  The middle panel shows a baseline type with a moderate amount of contamination, including several delay bins where many of the baselines are found in the lower right quadrant.  Lastly, the bottom panel shows the worst case.  Here, significant fractions of points within a delay bin (often well over 50\%) extend far into the lower right quadrant.  In the first two cases, the measured data are largely consistent with noise and contamination with the common mode systematic contributes a small fraction of outliers.  On the worst baseline type shown in Figure \ref{fig:phasevar} (28\,m east-west), however, the data are not distributed like noise. The systematic contamination cannot be considered an outlier.  As we discuss in \S\ref{sec:flagging_results}, baselines types that show this level of systematic contamination should be discarded entirely.

\subsubsection{Flagging Results}
\label{sec:flagging_results}

\begin{figure}
    \centering
    \includegraphics[width=\columnwidth]{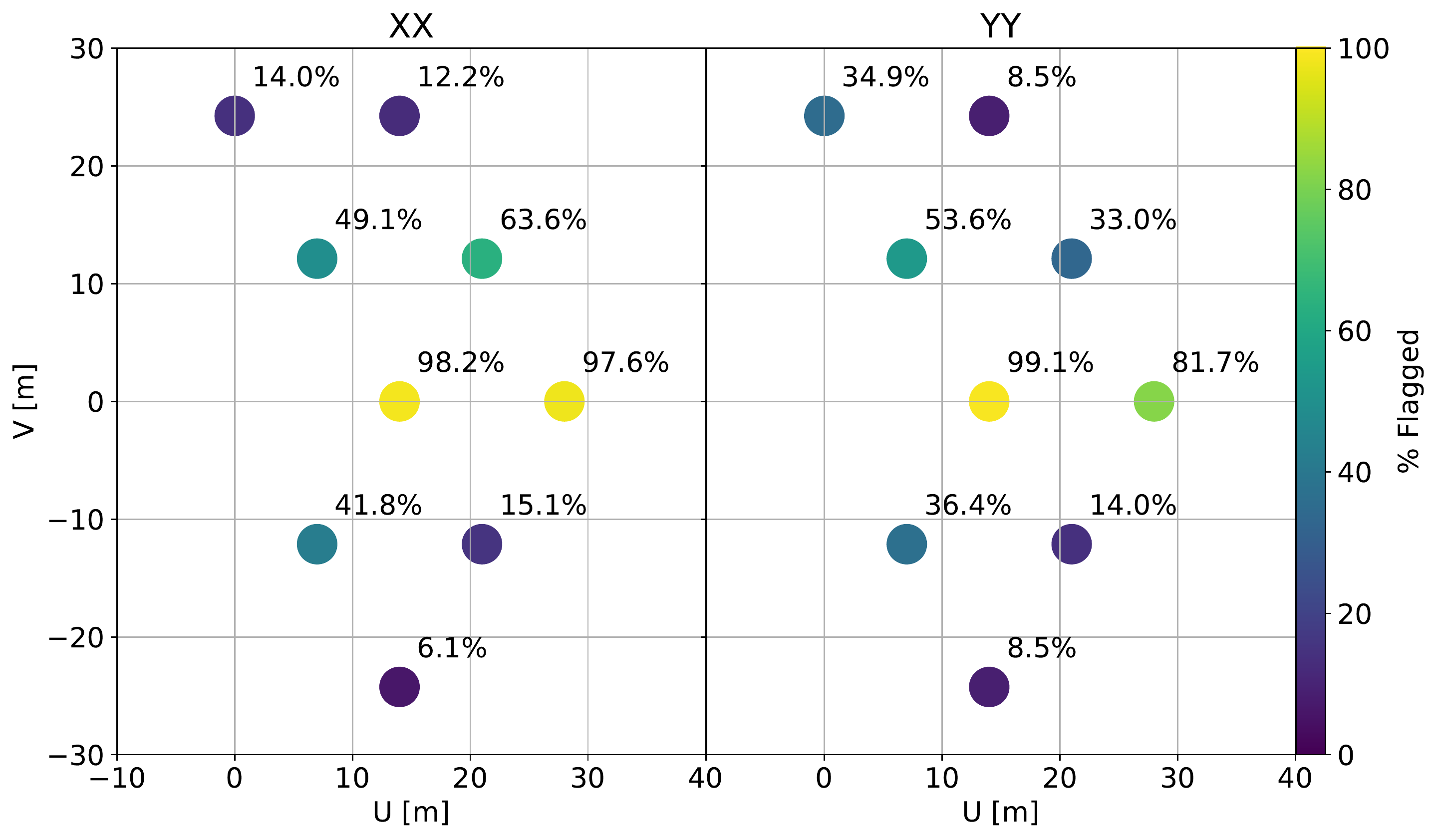}
    \caption{Fraction of data removed by our common mode flagging for each baseline type (indicated by its position in $uv$ space) and polarization. For the final analysis, we discard all data from both east-west baseline types in both polarizations.}
    \label{fig:flags_vs_uv}
\end{figure}

So far, we have attempted to detect the common mode systematic on a per-delay mode basis. However, flagging individual delay modes would cause every power spectrum $k$ mode to have a different error bar. We avoid this added complexity by extending flags across most of the delay spectrum.

In our Figure \ref{fig:systematic} example, the systematic only appears on positive delays. In general, approximately 50\% of baselines showing common mode contamination are only affected in either positive or negative delay modes.  Therefore, in order to avoid over-flagging, we flag \emph{all} delay spectrum points on that baseline having the same delay sign.  Baselines that show a common mode signal in both positive and negative delay bins are flagged entirely.

Figure \ref{fig:flags_vs_uv} shows how much data is flagged using this approach as the percentage of baselines flagged plotted in $uv$ space (which measures the positional baseline vector). Both the colors of each point and the percentages labeling them show the amount of total data discarded from each baseline type.  As can be seen, the common mode signal most affects the two east-west baseline types, and we discard all data from these two orientations from our final analysis.  On the non-east-west baselines, the amount of flagging varies from $\sim5\%$ to $\sim65\%$.  In total, we discard nearly 50\% of the data, which breaks down as 100\% of the data on the two east-west baseline types and $\sim30\%$ of the data on the remaining seven other types.


\subsubsection{Discussion}
\label{sec:systematic_discussion}

Since this paper provides the first description of this systematic it is worthwhile to provide a brief discussion of the physical nature of the effect. The systematic is best described as ``an excess correlation which does not change with the sky, occurring on delays much larger than those expected due to sky geometry.'' Such signals are often generically called ``common mode''.

Its distribution (Figure \ref{fig:flags_vs_uv}) shows similar behavior on both polarizations.  The fraction of flagged baselines appears roughly linear with baseline length, with $\sim$75\% of the 14\,m baselines flagged, $\sim$50\% of the 24.24\,m baselines flagged, and $\sim$40\% of 28\,m baselines flagged. However, there are only three baseline lengths represented in the data set analyzed here, so further work is needed to establish the robustness of this trend. The signal appears at delays more than twice as long as the baseline length but much shorter than the coarse band harmonics. It very often appears on only one side of the delay spectrum. 

Such signals have been observed in other instruments. \citet{kern_et_al_2019,kern_et_al_2020a} describe similar effects in HERA. One important difference for the HERA systematics, however, was that the high delay signals were observed to exhibit phase rotation versus time like the sky.  Later investigation using the phase to do rudimentary direction finding pointed to re-radiation of sky signals from a cable junction back to other nearby antennas \citep{heramemo104}.  The common mode observed in MWA data does not seem to evolve like the sky, but with only 30 minutes of LST, this conclusion could be strengthened with more data. However, the information we have at present suggests the common mode signal seen here is not the type of cross-talk as seen in the HERA Phase I system.

\begin{figure}
\centering
\includegraphics[width=0.8\columnwidth]{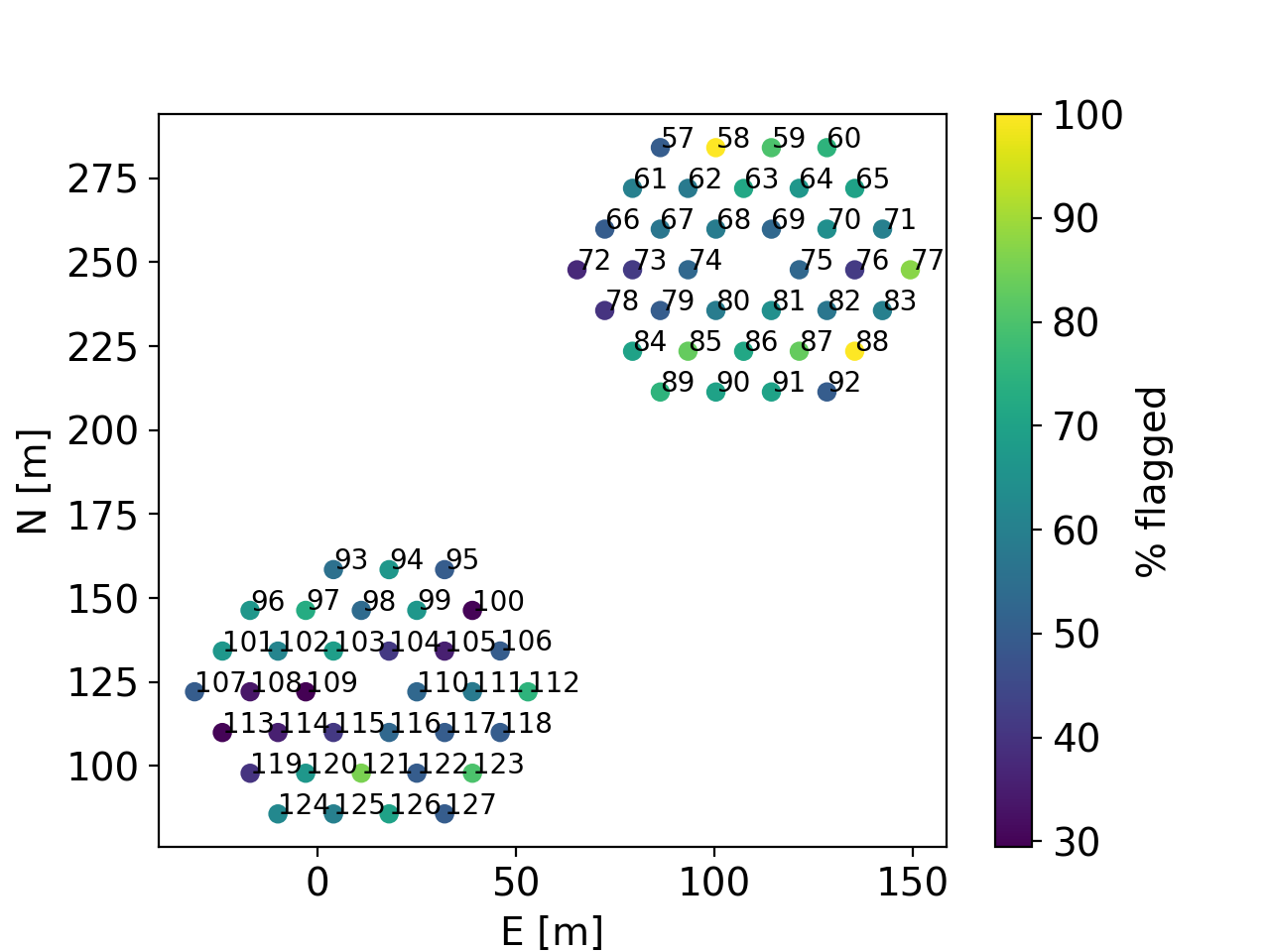}
\caption{A map illustrating how often each antenna is part of a baseline flagged for the high delay common mode. Though the distribution is suggestive of localized sources, the geographic pattern is far from decisive.}
\label{fig:baseline_map}
\end{figure}

One potential mechanism that does not involve cross-talk is a broadband signal radiating from a single source and reflecting at the receiver junction to acquire a differential delay of between 200 to 400\,ns.  This ``single radiator'' theory would show strong dependence on location within the array in both amplitude and phase.  Indeed we do see some intriguing evidence of spatial dependence when we investigate how often each antenna participates in a baseline that shows the common mode systematic, as in Figure \ref{fig:baseline_map}. 
Clusters of antennas with little flagging can be found in both the southwest and northeast of the southern hexagon, while the western edge of the northern hexagon also appears relatively unaffected.  In the northern hexagon, antennas 58 and 88 also stand out as completely flagged.
However, there are enough outliers and other non-uniformities which undercut the promise of these clues.

Noise-wave radiation could provide another mechanism for excess correlation.  Low noise amplifiers produce noise at the input, which can sometimes be substantially louder than the noise figure. Radiating from the antenna, these could be picked up by nearby antennas and observed as excess correlation.  However, without any extra delay introduced this would should up as a common mode at delays equal to the light travel time across the length of the baseline (i.e. the horizon delay).  A mechanism would be needed to increase the delay significantly, such as reflections between the amplifier and antenna or a phase slope introduced by the antenna match.

Further work is needed to better understand this systematic, its physical origins as well as the implications for power spectrum analysis. The risk of underestimating foreground bias or, worse, missing a 21\,cm detection due to negative bias should be considered in future power spectrum analyses.

\section{Simulation}
\label{sec:sims}

The flagging of common mode contamination is the last step in our pre-processing analysis before PS estimation with \texttt{simpleDS}.  Before we describe PS estimation, however, we first present the details of an LST-matched foreground-only sky simulation.  This simulated observation accompanies the data through the PS estimation to inform uncertainty calculations and to understand the leakage of foregrounds due to flagging and other analysis choices.

Simulated observations were made with the  \texttt{pyuvsim} \citep{lanman_et_al_2019} python package\footnote{http://github.com/RadioAstronomySoftwareGroup/pyuvsim}. \texttt{pyusim} is a visibility calculator which performs the radio interferometer measurement equation (RIME) integral as an Euler sum without flat sky or flat array approximations. It is optimized to be a precision reference for validation of accelerated codes which make more approximations. The package has been validated to one part in 10,000 using multiple comparison points. Most recently added are tests which validate diffuse integration using flux patterns which have analytically or arbitrarily convergent solutions to the RIME \citep{lanman_et_al_2020}.

The sky input model was a summation of GLEAM all-sky catalog \citep{hurley-walker_et_al_2017}, the Global Sky Model \citep{deOliveira-Costa_et_al_2008}, and an addition of the bright ``A-Team'' sources missing from GLEAM \citep{FHDmemo9, byrne_et_al_2022}. Visibilities corresponding to this sky model were computed over the 30 minute observation sampled every 16\,s to match with the final integration time of the LST-binned data products.

This simulation was performed for only one baseline from each redundant group, as identified in Figure \ref{fig:bl_select}. This reduced the total run time of the simulation by an estimated factor of 100. These simulated visibilities were then copied to each baseline.

The inclusion of both GLEAM and GSM sources will result in some double counting of point sources. As such, we expect the total sky power to be higher in the simulation compared to observed data.  However the inclusion of the GSM is necessary to model the effects of the Galaxy observed in the sidelobes of the MWA beam at the horizon.  This model is not being used to subtract foregrounds or calibrate, so we judge this inaccuracy to have unimportant and generally conservative impacts on the analysis.  The comparison between the simulated visibilities and the data is performed in PS space and so is discussed along with the results in \S\ref{sec:discussion}.

\begin{figure*}
    \centering
    \includegraphics[width=.95\textwidth]{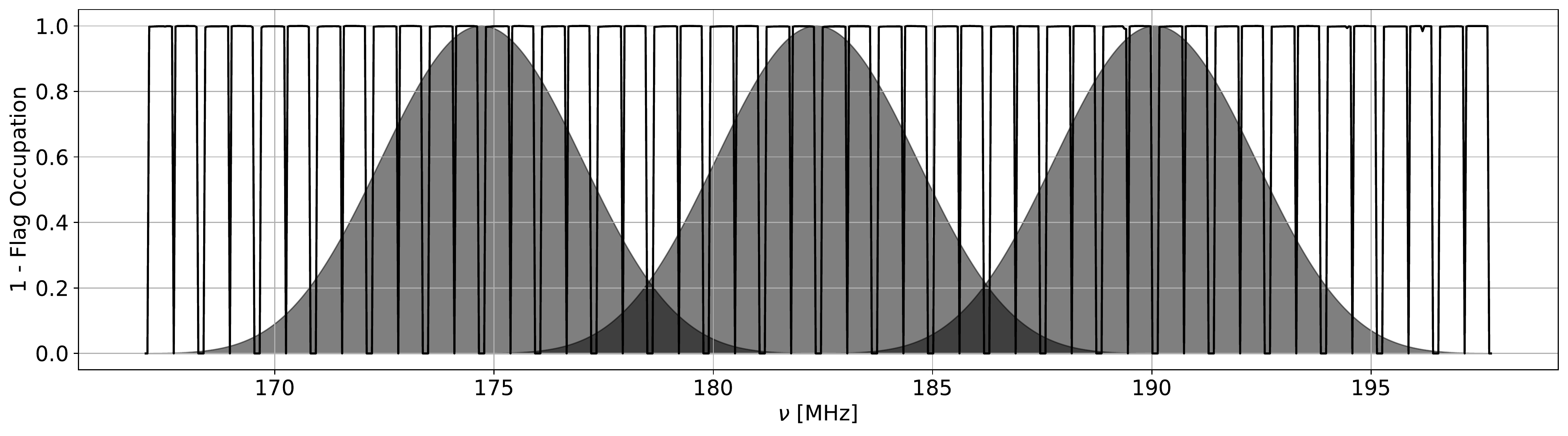}
    \caption{
    The Blackman-Harris tapered spectral windows selected in this analysis plotted over the average flagging occupation.
    }
    \label{fig:spectral_windows}
\end{figure*}

\section{simpleDS}
\label{sec:simpleDS}
Developed as an independent power spectrum analysis tool, \texttt{simpleDS} implements a variant of the delay spectrum method. It has been used to perform 21\,cm power spectrum estimation for PAPER \citep{kolopanis_et_al_2019}, and validating the HERA power spectrum pipeline \citep{aguirre_et_al_2022}. The complexity of interferometric 21\,cm pipelines can often make issues hard to diagnose or even obscure them entirely. The goal of \texttt{simpleDS} is to estimate a delay power spectrum with as readable code and in as few steps as possible. Its design focuses on simplicity and adherence to propagating units through all cosmological calculations and noise estimates by relying on definitions in \citet{liu_et_al_2014a} and \citet{HERAmemo113}.

The current analysis begins by selecting 3 spectral windows each 15.36\,MHz wide (half of the 30.72\,MHz total bandwidth) centered at 174.72\,MHz, 182.36\,MHz, and 190.07\,MHz. Each of these windows is carried through separately for the entire analysis. While the high and low frequency windows overlap with the middle window, a Blackman-Harris tapering function is applied to each window before the delay transform is calculated.  As shown in Figure~\ref{fig:spectral_windows}, the tapered windows have only minimal overlap and serve as functionally independent estimates of the PS at three different redshifts ($z$ = 7.14, 6.79, and 6.48 for the low, middle, and high frequencies windows, respectively).

Using the \texttt{simpleDS} package, all baselines are then Fourier Transformed from frequency to delay space.\footnote{As mentioned, each Blackman-Harris tapered spectral window is analyzed independently.  For brevity, this will no longer be explicitly stated, but can assumed in the following description.} The power spectrum is calculated by taking the outer product of all redundant baselines cross multiplying ``even'' and ``odd'' lst samples. This forms all baseline cross-products between the two independent time samples. Noise bias will be introduced if any measurements are squared. Wherever it occurs, multiplication of data by itself is avoided. Thus, only products between baseline A in the even set and baseline B in the odd set are included. Note that this amounts to a double avoidance of cross multiplication: between different baselines and between different time samples.

The average of these cross products forms the estimator for the delay power spectrum. The power spectrum is calculated for each polarization, baseline cross product within a baseline type, time,  delay mode (which includes both positive and negative delays), and across baseline type. In the final step these are averaged into a single power spectrum. On occasion we averaged only over a subset, for example inspecting the average of just a single baseline type. 

Following \citet{liu_et_al_2014a} and \citet{HERAmemo113}, the delay power spectrum estimator can be reinterpreted as an estimator for the cosmological power spectrum by applying the transformation 
\begin{equation}
    P(k_{\perp}, k_{\parallel}) = \Phi \left\langle  \tilde{V}_{i}(\tau) \tilde{V}_{j}(\tau) \right\rangle_{i\ne j}
    \label{eqn:power_spectrum}
\end{equation}
where 
\begin{equation}
    \Phi =  \left(10^{-23} \frac{c^{2}}{2k_{\rm B}} \right)^{2}\left[ \int d\nu\ \nu^{4} \frac{\Omega_{\rm pp}(\nu) \phi^{2}(\nu) }{ X^{2}(\nu) Y(\nu) } \right]^{-1}
    \label{eqn:normalization}
\end{equation}
where $i$ and $j$ denote independent baselines, $c$ is the speed of light, $k_{B}$ is the Boltzmann constant, $\phi(\nu)$ is the spectral tapering function, $X(\nu)$ and $Y(\nu)$ are cosmological unit conversion functions,  and
$\Omega_{\text{pp}} = \int dl dm \left| A(l, m) \right|^{2}$ is the integral of the square of the primary beam over the sky \citep{parsons_et_al_2014, parsons_2016}.  In this work, we use the \cite{sutinjo_et_al_2015} model for the MWA, as implemented in \texttt{pyuvdata} \citep{hazelton_et_al_2017}, to calculate the primary beam terms.

An average along the time dimension is performed and weighted by the expected thermal variance in each time bin; this variance is discussed in \S\ref{sec:thermal_noise}.
Then the positive and negative $k_{\parallel}$ modes are averaged together to form an estimator of $P(k_{\perp}, \left| k_{\parallel} \right|)$. To estimate the cosmological power spectrum as a function of wave number magnitude $P(|k|)$, redundant baseline groups of identical length but different orientations (triangles in Figure~\ref{fig:bl_select}) are averaged together weighted by their respective thermal variances. Finally, the two linear polarizations are also averaged together with the same weighting scheme to form the final cosmological power spectrum estimator.

 

\subsection{Uncertainty Estimates}
\label{sec:thermal_noise}

Uncertainty of the delay power spectrum is estimated using two methods: a theoretical noise calculation based on sky model and observing parameters, and bootstrap resampling over redundant baseline pairs. The strength of the theory calculation is in its relative simplicity, which allows more opportunities for verification. However, it does not account for non-thermal errors arising from instrument variation.  On the other hand, bootstrap resampling can provide an estimate of instrumental error but will under-estimate error if correlation between data points is not properly handled \citep{ parsons_et_al_2014,cheng_et_al_2018, tan_et_al_2021}. Here we report errors using both methods.  Lastly, as a consistency check, we compare both error estimates against a noise simulation processed through the pipeline.

As data are accumulated through repeat observing, some measurements at certain sidereal times become significantly better sampled than others. The variable number of data points accumulated, illustrated in the bottom right of Figure \ref{fig:lstbin}, drives the overall sensitivity. 
The presence of foreground power (either intrinsic to a given delay mode, or scattered there due to leakage) also increases the variances through foreground-noise cross terms when the data are squared.  (These cross terms have an expectation value of zero, but the foregrounds are bright enough that convergence to zero typically takes more independent samples then we have in practice.) 
Theoretical noise limits are calculated according to \citet{tan_et_al_2021} accounting for both sample count and variation due to sky signal (the so-called $P_S$ term, which comes from the simulations described in \S\ref{sec:sims}). However only the $P_N$ is used to weight the data during averaging. The detailed calculations are as follows.

The thermal noise estimation introduced in \citet{kolopanis_et_al_2019} has been updated using information in \citet{HERAmemo113}. A per frequency, per time integration noise amplitude is calculated using equations 2 and 3 from \citet{kolopanis_et_al_2019}:

\begin{equation}
    \sigma_{n}^{2} = \frac{T_{\text{sys}}(\nu)^{2}}{\Delta\nu t_{\text{int}} N_{\text{samples}}(\nu, t)} \label{eqn:sigma_noise}
\end{equation}
where the $\Delta\nu$ is the bandwidth between frequency bins, $t_{\text{int}}$ is the integration time of a time bin, $N_{\text{samples}}(\nu, t)$ is the total number of samples in a frequency-time bin accounting for LST averaging and is explicitly assumed to be a function of both time and frequency, and $T_{\text{sys}}(\nu)$ follows from \citet{rogers_bowman2008}:
\begin{equation}
    T_{\text{sys}}(\nu) = 180~K \left( \frac{\nu}{180~\text{MHz}}\right)^{-2.55} + T_{\text{rcvr}} \label{eqn:t_sys}
\end{equation}
We use a receiver temperature ($T_{\text{rcvr}}$) of 50\,K for an MWA tile \citep{tingay_et_al_2013a}.

Equation \ref{eqn:sigma_noise} is used to generate Gaussian random noise ($n_{i} \sim \mathcal{CN}(0, \sigma_{n})$) which is analyzed in parallel with the input data.

Additionally, a theoretical estimate of the expected thermal noise variance in the complex 21\,cm power spectrum\footnote{Specifically being the contribution to the complex 21\,cm power spectrum, the real and imaginary components contributing half the variance.} is computed  following section 3.2 in \citet{HERAmemo113}:
\begin{equation}
    P_{\text{N}}(t) = \text{Var}(\mathbf{P}) = \frac{1}{t_{\text{int}}} \frac{\int d\nu\ \nu^{4}\phi(\nu)^{2} \frac{T_{\text{sys}}(\nu)^{2}}{N_{\text{samples}}(\nu,t)} \Omega_{\text{p}}^{2} }{ \int d\nu\ \nu^4 \phi(\nu)^{2} X(\nu)^{-2}Y(\nu)^{-1}\Omega_{\text{pp}}} \label{eqn:p_n}
\end{equation}
where $\Omega_{\text{p}} = \int dl dm\ A(l, m)$ is the integral of the primary beam over the sky \citep{parsons_et_al_2014, parsons_2016}.

This integral provides an estimate of the thermal contribution to the 21cm power spectrum constant across all delays for each time bin and can be propagated through any additional averaging steps.

The total uncertainty on the 21\,cm power spectrum (assuming no covariance between the sky signal and noise terms) is a combination of the thermal estimate ($P_{\text{N}}$) and an estimated sky signal ($P_{\text{s}}(k)$) \citep{kolopanis_et_al_2019, tan_et_al_2021}:
\begin{equation}
    P_{\text{SN}}(k)^{2} = 2 P_{\text{S}}(k) P_{\text{N}} + P_{\text{N}}^{2} 
\end{equation}
where our simulation from \S\ref{sec:sims} is used as a proxy for $P_{\text{S}}$.


Bootstrap errors are calculated by sampling with replacement among all available cross-multiplied pairs of baselines 100 times, averaging in time, and taking the variance of these resamplings as an estimator of the variance of the delay power spectrum.  Here we re-sampled only across the redundant baseline axis and not in the time dimension. This avoids potential oversampling of correlated measurements.  We verified that when applied to a simulation with known noise and foreground power it reproduces the expected variance.

\section{Results}
\label{sec:results}
\begin{figure*}
    \centering
    \includegraphics[width=\textwidth]{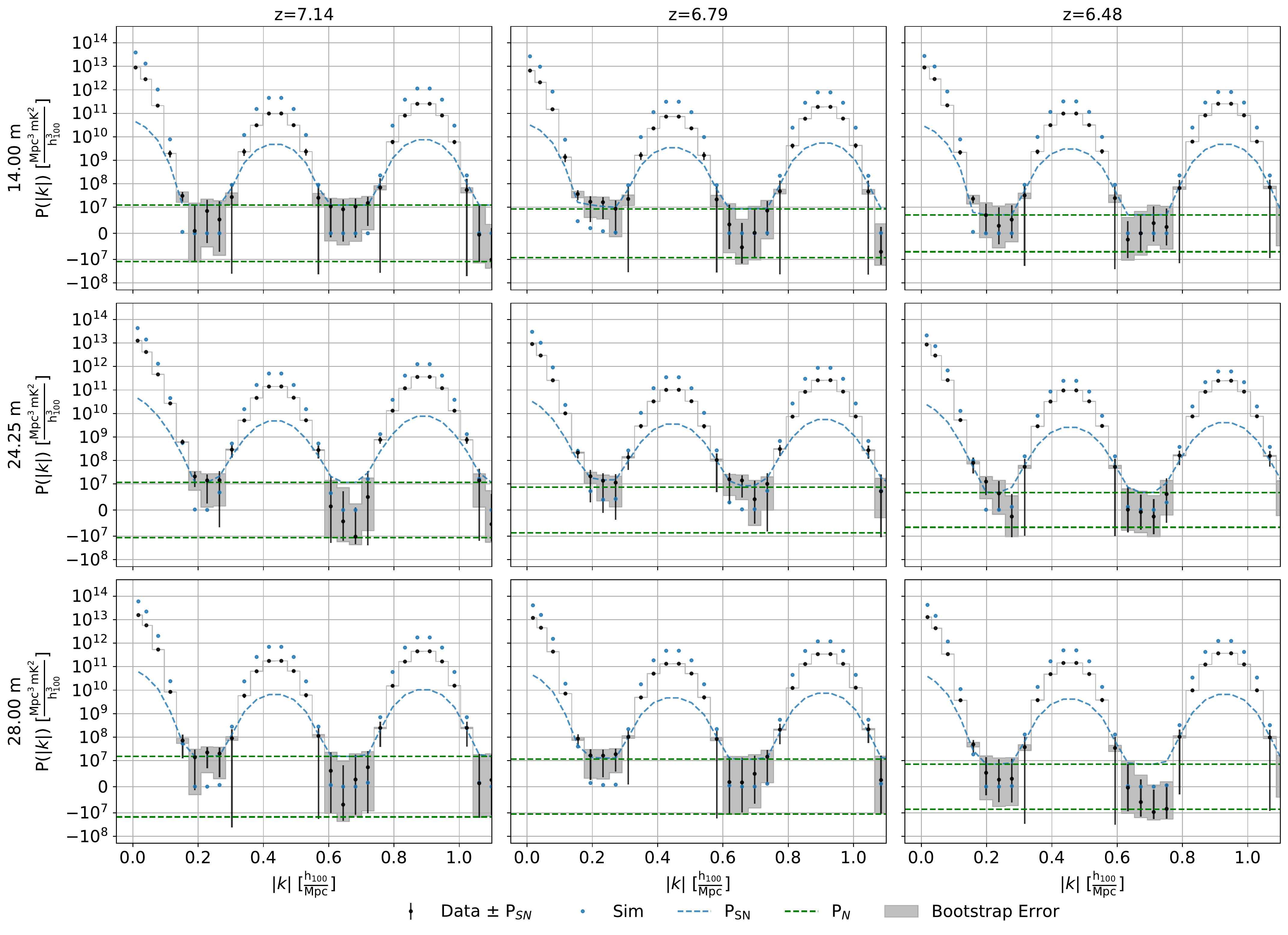}
    \caption{The power spectrum across three baseline lengths and redshift bins. All axes not indicated here have been averaged across.  Black points indicate the measured values, wile blue points are from our foreground simulations. Simulated values are a factor of 2 to 4 higher than observed. Possible explanations for this disagreement described in the text.  The blue-dashed and green-dashed lines show our calculations of $P_{\rm SN}$ and $P_{\rm N}$, respectively. The $y$ axis is a symmetric-log scale that highlights  the noise-limited bins.  The error bars resulting from theory (black lines) and from bootstrapping (gray shaded) show very good agreement in this regime here as well as at higher power levels, though these are less relevant.}
    \label{fig:ps_noise}
\end{figure*}

The results of our power spectrum estimation are shown in Figure \ref{fig:ps_noise}.
Each column shows one of the three redshifts from the sub-bands described in \S\ref{sec:simpleDS}.  Each row shows one of the three baseline lengths (recall that we have selected 9 redundant baseline types which can be grouped into three length bins).  Black points show the measured PS values and their error bars are 2$\sigma$ $P_{\text{SN}}$ uncertainties.  The blue points show the PS for our simulated visibilities described in \S\ref{sec:sims}.  The dashed green line shows the thermal noise level, $P_{\text{N}}$ (this estimate is used for any inverse variance weighting in time or across baseline type due to its stability across delay), and the dashed blue line shows $P_{\text{SN}}$.  The gray shaded regions show the 2$\sigma$ uncertainties estimated from bootstrapping.  We note that the bootstrapping and $P_{\text{SN}}$ methods for uncertainty estimation produce very similar results.  The $P_{\text{SN}}$ error bars, being simpler, less subject to potential loss, and more model driven are our preferred option. These are used to report as the consensus limit result.

The general features of this plot follow the usual pattern for the MWA.  At low values of $k$, the measurements are dominated by the intrinsic foregrounds and are well above the noise level.  The periodic structure of the coarse band edges in the MWA bandpass scatters this foreground power to higher $k$ values, resulting in the bands of contaminated PS values centered at $k\sim0.45$ and $k\sim0.9\, h_{100}\rm{Mpc}^{-1}$ (and subsequent harmonics of this scale at higher $k$ that are not plotted).  Between these ``coarse band peaks", however, the measurements are largely consistent with the expected noise level. We observe that the simulations are a factor of 2 to 5 higher than the data, but as noted, the simulation contains both diffuse and compact models without accounting for double counting of sources. This might account for some, but not all of this difference. As the foregrounds are not being used in a precision context like calibration we have elected to not pursue this difference at this time.

\begin{figure*}
    \centering
    \includegraphics[width=\textwidth]{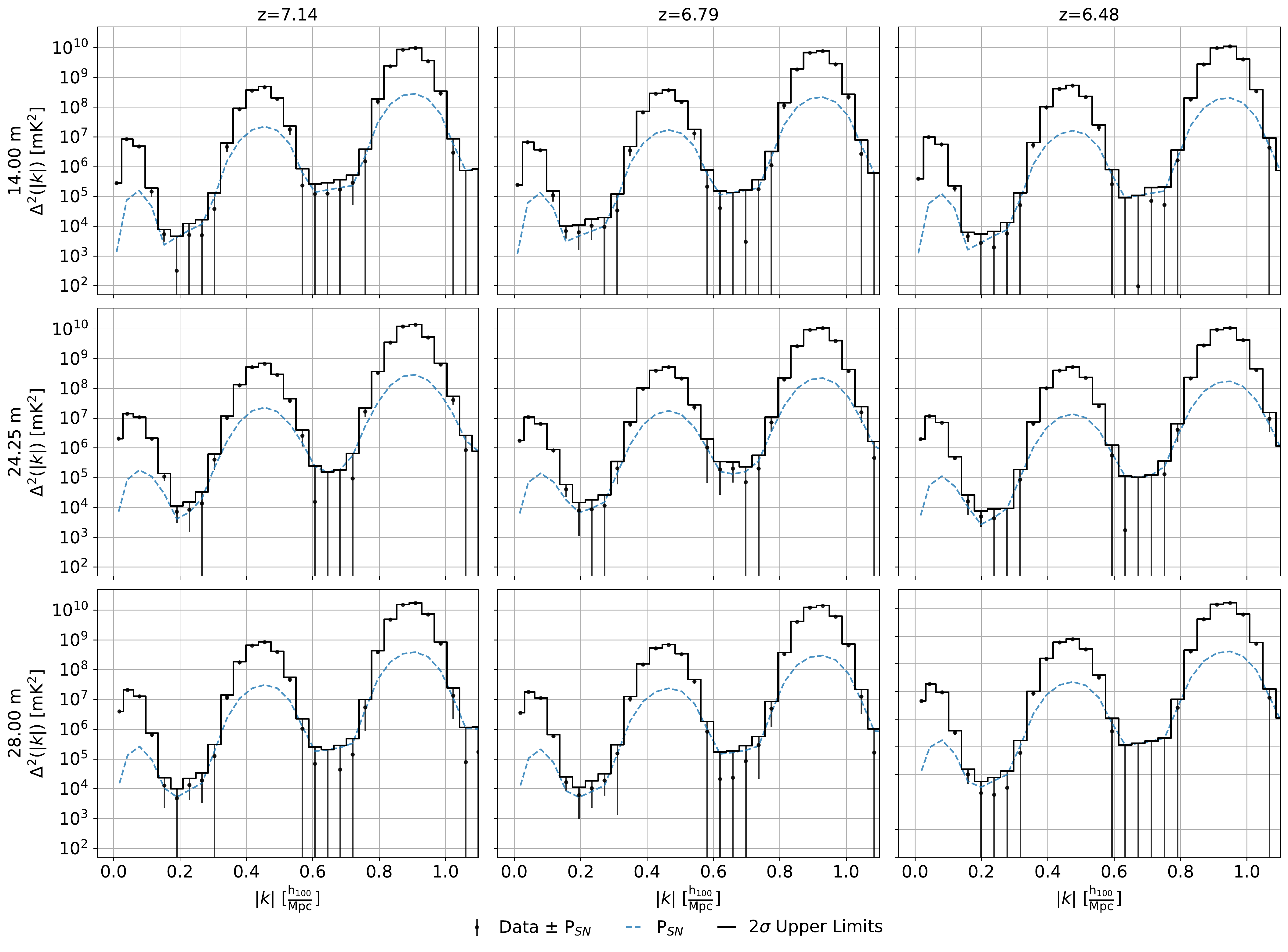}
    \caption{Same data as in Figure \ref{fig:ps_noise} but now converted to the dimensionless power spectrum ($\Delta^2$). Only the measured values (black points), $P_{\rm SN}$ (blue dashed line), $P_{\rm SN}$ error bars (thin black lines), and the $2\sigma P_{\rm SN}$ upper limits (thick black line) are shown.  Between the coarse band peaks, many measurements are consistent with zero.}
    \label{fig:ps_lims}
\end{figure*}

Figure \ref{fig:ps_lims} shows the measurements and 2$\sigma\, P_{\text{SN}}$ uncertainties from Figure \ref{fig:ps_noise} plotted as the dimensionless power spectrum $\Delta^2 = \frac{k^3}{2\pi^2}P(k)$.  The blue dashed line again shows $P_{\text{SN}}$, while the thick black line shows the 2$\sigma$ upper limits on the 21\,cm power spectrum.  As in \citet{hera_collaboration_2022}, in calculating our final 2$\sigma$ upper limits, we use the $2\sigma$ uncertainty above 0 for any point where the measured power spectrum value is negative.  All limit values are reported in Appendix \ref{appendix:tables}.  We find a lowest limit of $4.58 \times 10^{3}$ [mK]$^{2}$ at $k=.190 h_{100}Mpc^{-1}$ and a redshift of z=7.14. The lowest values for each redshift bin and baseline group are shown in Table \ref{tab:upper_lims}.

\begin{table*}
    \centering
    \begin{tabular}{ccccccc}
     & \multicolumn{2}{c}{z=7.14}& \multicolumn{2}{c}{z=6.79}& \multicolumn{2}{c}{z=6.48}\\
     $|\textbf{b}|$ [m] & k $\left[\dfrac{h_{100}}{\texttt{Mpc}}\right]$& $\Delta_{\text{UL}}^{2}$ [mK$^{2}$]  & k $\left[\dfrac{h_{100}}{\texttt{Mpc}}\right]$& $\Delta_{\text{UL}}^{2}$ [mK$^{2}$]  & k $\left[\dfrac{h_{100}}{\texttt{Mpc}}\right]$& $\Delta_{\text{UL}}^{2}$ [mK$^{2}$] \\\hline
     14 & 0.190  &  4.58 $\times 10^{3}$ & 0.155 &  9.96 $\times 10^{3}$ & 0.198 &  5.52 $\times 10^{3}$\\
     24.25 & 0.190  &  1.13 $\times 10^{4}$ & 0.194  &  1.46 $\times 10^{4}$ & 0.198  &  7.64 $\times 10^{3}$ \\
     28 & 0.190 &  1.01 $\times 10^{4}$ & 0.194  &  1.14 $\times 10^{4}$ & 0.199  &  5.55 $\times 10^{3}$
    \end{tabular}
    \caption{The lowest upper limit values and their associated wavenumber on the dimensionless 21cm power spectrum from each baseline type in each redshift bin from this work. These correspond to the lowest value of the solid black line in each sub panel of Figure~\ref{fig:ps_lims}.}
    \label{tab:upper_lims}
\end{table*}




\section{Discussion}
\label{sec:discussion}

\begin{figure*}
    \centering
    \includegraphics[width=\textwidth]{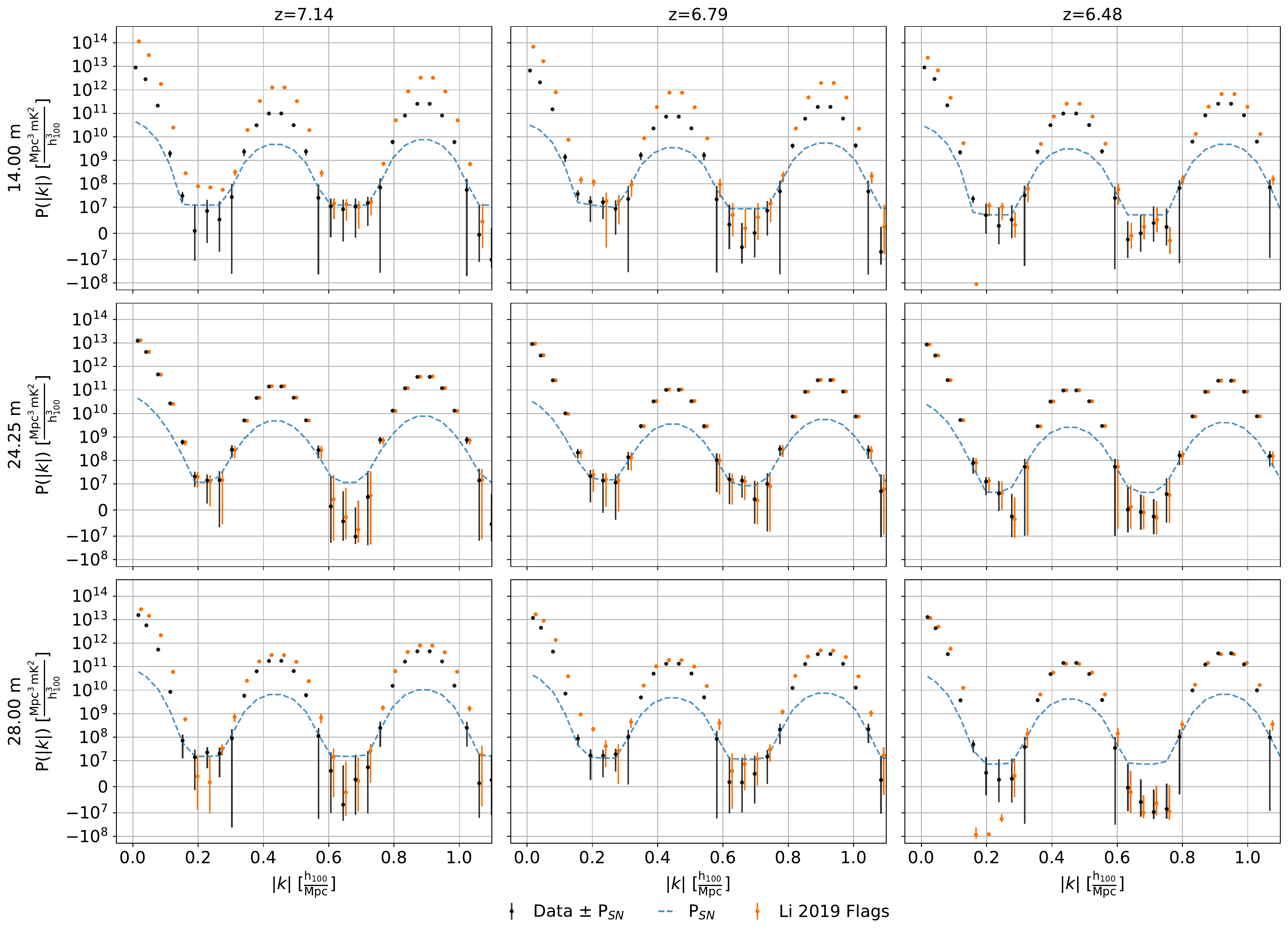}
    \caption{Effect of common mode flagging.  Black points show the results of the present analysis, while orange points (slightly offset in $k$ for visual clarity) use only the flags from the \citet{li_et_al_2019} analysis, which did not address the common mode systematic.  Without common mode flagging, we see several points in redshift bin 6.48 which are significantly negative; after flagging for common mode, these points are now conistent with zero. Common mode flagging also reduced the overall foreground power, most notably on the 14\,m baseline (see text for explanation). Finally, note the 28\,m redshift 7.14 bin where several points went from a null result (i.e. consistent with zero to within the noise) to a statistically significant detection. This could be evidence for negative systematics suppressing real positive bias.  The removal of negative systematics allowed the true positive bias to be seen.}
    \label{fig:ps_flags}
\end{figure*}

One of the main results of this analysis is the impact of the common mode systematic flagging on the final PS. This is  illustrated in Figure \ref{fig:ps_flags} which shows the power spectrum before and after common mode flagging.  
The orange points and error bars (offset in $k$ from the black points for visual clarity) show the PS measurements and $2\sigma\, P_{\text{SN}}$ uncertainties we obtain if we use only the original data flags from \citet{li_et_al_2019}, i.e., if we omit the additional flagging described in \S\ref{sec:quality2}.  

There are several interesting features that can be seen.  First, the additional flagging appears to significantly decrease the measured foreground power (particularly on the 14\,m baselines in the top row).  This is an expected result, as our additional flagging removes all 14\,m east-west baselines (c.f. Figure \ref{fig:flags_vs_uv}), leaving only the baselines oriented towards the NW or SW.  Due to a somewhat chance alignment of the fringe pattern of an east-west baseline with the Galactic plane, these baselines do in fact see significantly more power than their 14\,m north-east and south-east counterparts, which we confirm with our simulations.  Removing this excess power reduces foreground contamination, as can be seen most prominently around $k\sim0.2\, h_{100}\rm{Mpc}^{-1}$ at $z=7.14$ for the 14\,m baselines (i.e. in the top left corner).  However, we emphasize that we did not flag the east-west baselines because of their increased foreground power, but because of significant common mode contamination.  (Whether the increased foreground power and the increased common mode contamination are related remains an open question.)

A second key feature is the elimination of highly negative power spectrum points, as can be most clearly seen in the 14\,m and 28\,m baselines near $k\sim0.2\, h_{100}\rm{Mpc}^{-1}$ at $z=6.48$ (i.e. the top and bottom right plots).  With only the original \citet{li_et_al_2019} flags applied, the common mode signals remain in the data and are often out of phase on nominally redundant baselines (i.e. the phase of the common mode signal on one baseline is $\sim180^\circ$ off from a common mode signal on a baseline of the same length and orientation).  As explained above, the net effect of cross-multiplying these baselines is to produce a highly significant negative power spectrum value.  Such a measurement is obviously unphysical for sky or noise signals, but nevertheless is the result of failing to remove the common mode systematic.

Most interestingly, perhaps, is the effect of our additional flagging on the points near near $k\sim0.2\, h_{100}\rm{Mpc}^{-1}$ at $z=7.14$ on the 28\,m baselines (i.e. the lower left plot).  Taken at face value, the orange points in this area look like noise-dominated measurements that would place relatively competitive limits on the 21\,cm PS signal strength.  However, once the common mode is flagged, those points become statistically significant detections of residual foreground power.  The explanation is again negative measurements resulting from cross-multiplication of out-of-phase common mode signals; however, in this case, the signals are only present on some baselines, and when all the 28\,m baseline power spectra are averaged together, the negative powers serve to pull the average back towards zero but are not numerous enough to cause a significantly negative PS measurement.  This is in fact the most pernicious presentation of this systematic error.  The negative points in the $z=6.48$ band are obviously unphysical and motivated a reanalysis of this data.  Had we only been presented with the $z=7.14$ power spectra, however, we might have concluded they were reasonable measurements and incorrectly interpreted them as strong constraints on the 21\,cm PS amplitude.  This serves to illustrate that the practice of treating systematics in 21\,cm PS measurements as purely positive biases (e.g. as in the likelihood used to marginalize over systematic offsets in the first round of HERA results) is insufficient \citep{hera_h1c_theory}.  Negative systematics are also possible, which means the nominally conservative practice of treating all measurements as upper limits leaves room for error if systematics have not been sufficiently mitigated.

It is worthwhile to consider whether there are variants of the ``standard'' delay spectrum technique used here that could mitigate or prevent this bias.  In an analysis motivated by the results of this present work, \citet{morales_et_al_2022} construct a toy model of a baseline-dependent systematic and suggest that including the ``auto-baseline'' terms (i.e. cross-multiplying the same baseline at different, but closely spaced, times) can prevent any negative bias or signal loss.  Similarly, one might imagine averaging together all the nominally redundant visibilities and then cross-multiplying only between times, which should have a similar effect.  \citet{tan_et_al_2021}, however, show that the probability distributions for noise-dominated measurements can become skewed in these cases, complicating statistical interpretation of upper limits.  Cross-multiplying over both axes (time and baseline, as we do here) has the benefit of producing symmetric PDFs, but given the significant effect of systematics on our analysis, a skewed PDF might be a price worth paying.

\begin{figure*}
    \centering
    \includegraphics[width=\textwidth]{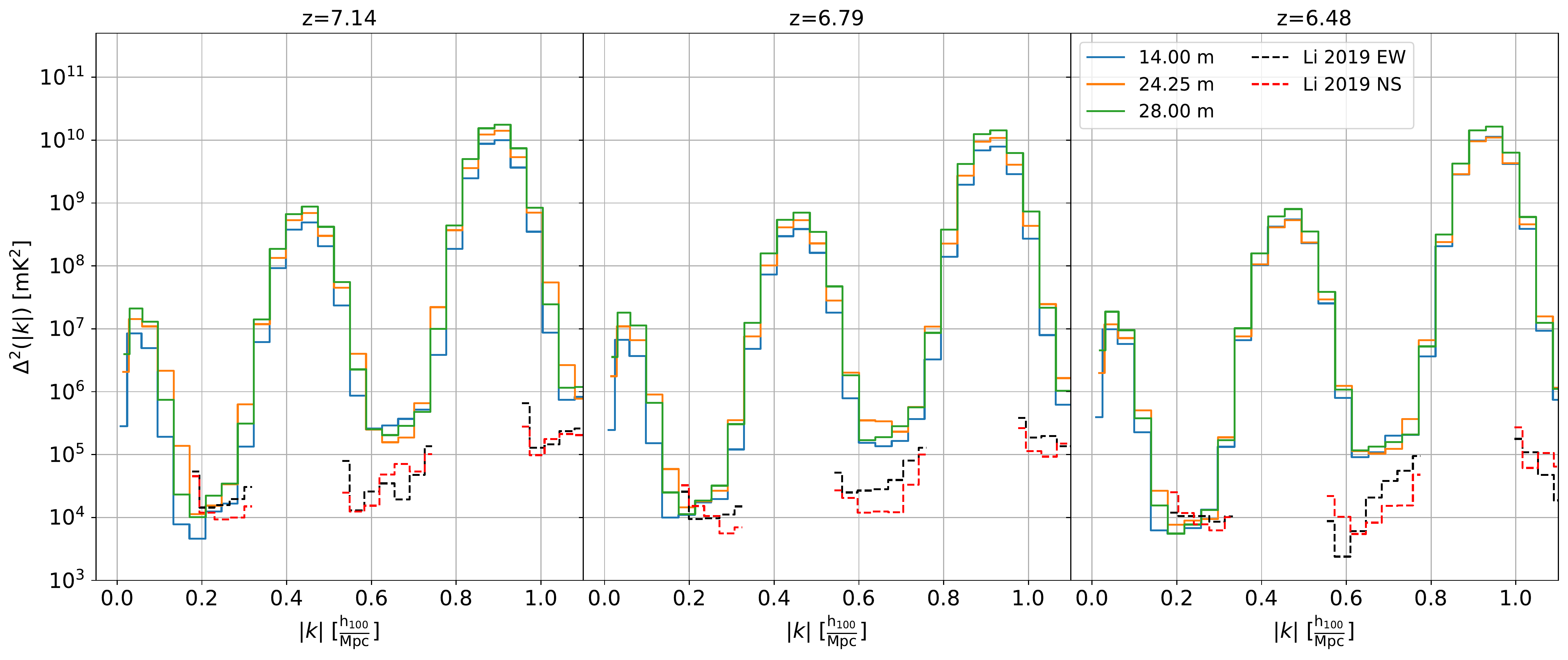}
    \caption{Comparison of our PS upper limits (blue, orange and green solid lines for the 14\,m, 24.24\,m, and 28\,m baselines lengths, respectively) with those from \citet{li_et_al_2019} (black and red dashed lines for the EW and NS polarizations, respectively).  In general, the \citet{li_et_al_2019} limits are lower given the increased sensitivity of that analysis (from using more observations and baselines). However, near$k\sim0.2\, h_{100}\rm{Mpc}^{-1}$, which is right at the edge of the foreground wedge, our limits are  lower, suggesting that our analysis is doing slightly better at mitigating foreground leakage.}
    \label{fig:ps_li}
\end{figure*}

It is also interesting to compare our final PS limits with those obtained via the FHD/$\varepsilon$ppsilon analysis of this data set from \citet{li_et_al_2019}, which we do in Figure \ref{fig:ps_li}.  For compactness, we plot the PS derived from each baseline length in the same panel (blue, orange, and green solid lines), as well as the limits from both polarizations of \citet{li_et_al_2019} (black and red dashed lines).  Note that while FHD/$\varepsilon$ppsilon naturally combines measurements from all baseline lengths, they choose not to combine polarizations as we do here.  As discussed above, although our data come from the same set of observations, there is only moderate overlap between the data contributing to the final PS with the main difference being in baseline length.  
They also use five different pointings of the MWA beam (as opposed to just one in our analysis), giving their analysis significantly lower thermal noise levels than we have here.  This fact is clearly noticeable between the coarse band harmonics (e.g. at $k\sim0.65\, h_{100}\rm{Mpc}^{-1}$ where, at $z=6.48$, \citealt{li_et_al_2019} obtain their lowest limit).  Here our limits are noise limited indicating the potential for further gains with the addition of more measurements. 

\citet{li_et_al_2019} also flag the coarse band contamination, leading to gaps in their PS versus $k$.  However, it is clear that the coarse band contamination is significantly wider in our current analysis (i.e. applying the same $k$ mask as \citet{li_et_al_2019} would not remove the full extent of the coarse band signatures from our PS limits).  A major contributor to this fact is the application of \emph{foreground subtraction} in the \citet{li_et_al_2019} analysis, which reduces the overall power that can scatter to high $k$.  Interestingly, however, near the intrinsic foreground contamination at $k\sim0.2\, h_{100}\rm{Mpc}^{-1}$, our limits are lower than those from \citet{li_et_al_2019}, suggesting that altogether, the delay spectrum approach can better limit ``bleed" into the EoR window (as suggested by \citealt{morales_et_al_2019}).  In fact, the best limits from this analysis come from this range ($k=0.190\, h_{100}\rm{Mpc}^{-1}$ in the $z=7.14$ band), as opposed to the best limit in \citet{li_et_al_2019}, which comes from much higher $k$ ($k=0.591\, h_{100}\rm{Mpc}^{-1}$ in the $z=6.48$ band) and is altogether only a factor of $\sim2$ lower than our best limit.

Lastly, it is interesting to consider how the presence of unflagged common mode systematics affected the \citet{li_et_al_2019} analysis.  The answer is not so straightforward.  In a delay spectrum analysis, individual baselines are cross-multiplied where the the product of two out of phase signals can lead directly to a negative value.  In FHD/$\varepsilon$ppsilon, individual visibilities are summed into the $uv$ plane during the gridding step; only after a full $(u,v,f)$ cube is built up and Fourier transformed to $(u,v,\eta)$ are the data squared. 
 In this approach, one might expect out-of-phase systematics to lead to signal loss --- i.e. adding them together will drive the sum towards zero --- but, given the presence of the squared terms (baseline A times baseline A) it seems more difficult to get a actual negative value.  Quantifying the magnitude of this effect will require an investigation in simulation.  While flagging these common mode systematics should become a standard piece of any MWA Phase II analysis, determining their exact impact on a reconstructed sky analysis remains an area needing more study.

\section*{Acknowledgements}

Support for MJK and DCJ has been provided by the National Science Foundation under grant numbers 2144995 and 2104350.  Support for JCP has been provided by the National Science Foundation under grant number 2106510.  SM received support from Brown University's Karen T. Romer Undergraduate Teaching and Research Awards (UTRA) program and the NASA Rhode Island Space Grant.  The authors would also like to thank Miguel Morales for helpful feedback and discussion.

This scientific work makes use of the Murchison Radio-astronomy Observatory, operated by CSIRO. We acknowledge the Wajarri Yamatji people as the traditional owners of the Observatory site. Support for the operation of the MWA is provided by the Australian Government (NCRIS), under a contract to Curtin University administered by Astronomy Australia Limited. We acknowledge the Pawsey Supercomputing Centre which is supported by the Western Australian and Australian Governments. The MWA Phase II upgrade project was supported by Australian Research Council LIEF grant LE160100031 and the Dunlap Institute for Astronomy and Astrophysics at the University of Toronto. This research was conducted using computation resources and services at the Center for Computation and Visualization, Brown University.

\section*{Data Availability}

The MWA data used in this project are available to the public via the All-Sky Virtual Observatory (\url{https://asvo.mwatelescope.org/}) under project code G0009. Derived data products are available from the authors on request.



\bibliographystyle{mnras}
\bibliography{mwa_phase2_delay_systematic} 

\begin{thebibliography}{}
\makeatletter
\relax
\def\mn@urlcharsother{\let\do\@makeother \do\$\do\&\do\#\do\^\do\_\do\%\do\~}
\def\mn@doi{\begingroup\mn@urlcharsother \@ifnextchar [ {\mn@doi@}
  {\mn@doi@[]}}
\def\mn@doi@[#1]#2{\def\@tempa{#1}\ifx\@tempa\@empty \href
  {http://dx.doi.org/#2} {doi:#2}\else \href {http://dx.doi.org/#2} {#1}\fi
  \endgroup}
\def\mn@eprint#1#2{\mn@eprint@#1:#2::\@nil}
\def\mn@eprint@arXiv#1{\href {http://arxiv.org/abs/#1} {{\tt arXiv:#1}}}
\def\mn@eprint@dblp#1{\href {http://dblp.uni-trier.de/rec/bibtex/#1.xml}
  {dblp:#1}}
\def\mn@eprint@#1:#2:#3:#4\@nil{\def\@tempa {#1}\def\@tempb {#2}\def\@tempc
  {#3}\ifx \@tempc \@empty \let \@tempc \@tempb \let \@tempb \@tempa \fi \ifx
  \@tempb \@empty \def\@tempb {arXiv}\fi \@ifundefined
  {mn@eprint@\@tempb}{\@tempb:\@tempc}{\expandafter \expandafter \csname
  mn@eprint@\@tempb\endcsname \expandafter{\@tempc}}}

\bibitem[\protect\citeauthoryear{{Aguirre} et~al.,}{{Aguirre}
  et~al.}{2022}]{aguirre_et_al_2022}
{Aguirre} J.~E.,  et~al., 2022, \mn@doi [\apj] {10.3847/1538-4357/ac32cd},
  \href {https://ui.adsabs.harvard.edu/abs/2022ApJ...924...85A} {924, 85}

\bibitem[\protect\citeauthoryear{{Barry}, {Beardsley}, {Byrne}, {Hazelton},
  {Morales}, {Pober}  \& {Sullivan}}{{Barry} et~al.}{2019}]{barry_et_al_2019a}
{Barry} N.,  {Beardsley} A.~P.,  {Byrne} R.,  {Hazelton} B.,  {Morales} M.~F.,
  {Pober} J.~C.,   {Sullivan} I.,  2019, \mn@doi [\pasa]
  {10.1017/pasa.2019.21}, \href
  {https://ui.adsabs.harvard.edu/abs/2019PASA...36...26B} {36, e026}

\bibitem[\protect\citeauthoryear{{Bowman}, {Rogers}, {Monsalve}, {Mozdzen}  \&
  {Mahesh}}{{Bowman} et~al.}{2018}]{bowman_et_al_2018}
{Bowman} J.~D.,  {Rogers} A. E.~E.,  {Monsalve} R.~A.,  {Mozdzen} T.~J.,
  {Mahesh} N.,  2018, \mn@doi [\nat] {10.1038/nature25792}, \href
  {https://ui.adsabs.harvard.edu/#abs/2018Natur.555...67B} {555, 67}

\bibitem[\protect\citeauthoryear{{Byrne}}{{Byrne}}{2021a}]{byrne_thesis}
{Byrne} R.,  2021a, PhD thesis, University of Washington, Seattle

\bibitem[\protect\citeauthoryear{{Byrne}}{{Byrne}}{2021b}]{FHDmemo9}
{Byrne} R.,  2021b, EoR Imaging Memo \#09

\bibitem[\protect\citeauthoryear{{Byrne}, {Morales}, {Hazelton}, {Sullivan},
  {Barry}, {Lynch}, {Line}  \& {Jacobs}}{{Byrne}
  et~al.}{2022}]{byrne_et_al_2022}
{Byrne} R.,  {Morales} M.~F.,  {Hazelton} B.,  {Sullivan} I.,  {Barry} N.,
  {Lynch} C.,  {Line} J. L.~B.,   {Jacobs} D.~C.,  2022, \mn@doi [\mnras]
  {10.1093/mnras/stab3276}, \href
  {https://ui.adsabs.harvard.edu/abs/2022MNRAS.510.2011B} {510, 2011}

\bibitem[\protect\citeauthoryear{{Cheng} et~al.,}{{Cheng}
  et~al.}{2018}]{cheng_et_al_2018}
{Cheng} C.,  et~al., 2018, \mn@doi [\apj] {10.3847/1538-4357/aae833}, \href
  {https://ui.adsabs.harvard.edu/abs/2018ApJ...868...26C} {868, 26}

\bibitem[\protect\citeauthoryear{{DeBoer} et~al.,}{{DeBoer}
  et~al.}{2017}]{deboer_et_al_2017}
{DeBoer} D.~R.,  et~al., 2017, \mn@doi [\pasp]
  {10.1088/1538-3873/129/974/045001}, \href
  {http://adsabs.harvard.edu/abs/2017PASP..129d5001D} {129, 045001}

\bibitem[\protect\citeauthoryear{{Dillon}}{{Dillon}}{2021}]{heramemo104}
{Dillon} J.,  2021, HERA Memo 104

\bibitem[\protect\citeauthoryear{{Eastwood} et~al.,}{{Eastwood}
  et~al.}{2019}]{eastwood_et_al_2019}
{Eastwood} M.~W.,  et~al., 2019, \mn@doi [\aj] {10.3847/1538-3881/ab2629},
  \href {https://ui.adsabs.harvard.edu/abs/2019AJ....158...84E} {158, 84}

\bibitem[\protect\citeauthoryear{{HERA Collaboration} et~al.,}{{HERA
  Collaboration} et~al.}{2022a}]{hera_h1c_theory}
{HERA Collaboration} et~al., 2022a, \mn@doi [\apj] {10.3847/1538-4357/ac2ffc},
  \href {https://ui.adsabs.harvard.edu/abs/2022ApJ...924...51A} {924, 51}

\bibitem[\protect\citeauthoryear{{HERA Collaboration} et~al.,}{{HERA
  Collaboration} et~al.}{2022b}]{hera_collaboration_2022}
{HERA Collaboration} et~al., 2022b, \mn@doi [\apj] {10.3847/1538-4357/ac1c78},
  \href {https://ui.adsabs.harvard.edu/abs/2022ApJ...925..221A} {925, 221}

\bibitem[\protect\citeauthoryear{Hazelton, Jacobs, Pober  \&
  Beardsley}{Hazelton et~al.}{2017}]{hazelton_et_al_2017}
Hazelton B.~J.,  Jacobs D.~C.,  Pober J.~C.,   Beardsley A.~P.,  2017, \mn@doi
  [The Journal of Open Source Software] {10.21105/joss.00140}, 2

\bibitem[\protect\citeauthoryear{{Hurley-Walker} et~al.,}{{Hurley-Walker}
  et~al.}{2017}]{hurley-walker_et_al_2017}
{Hurley-Walker} N.,  et~al., 2017, \mn@doi [\mnras] {10.1093/mnras/stw2337},
  \href {http://adsabs.harvard.edu/abs/2017MNRAS.464.1146H} {464, 1146}

\bibitem[\protect\citeauthoryear{{Kern}, {Parsons}, {Dillon}, {Lanman},
  {Fagnoni}  \& {de Lera Acedo}}{{Kern} et~al.}{2019}]{kern_et_al_2019}
{Kern} N.~S.,  {Parsons} A.~R.,  {Dillon} J.~S.,  {Lanman} A.~E.,  {Fagnoni}
  N.,   {de Lera Acedo} E.,  2019, \mn@doi [\apj] {10.3847/1538-4357/ab3e73},
  \href {https://ui.adsabs.harvard.edu/abs/2019ApJ...884..105K} {884, 105}

\bibitem[\protect\citeauthoryear{{Kern} et~al.,}{{Kern}
  et~al.}{2020}]{kern_et_al_2020a}
{Kern} N.~S.,  et~al., 2020, \mn@doi [\apj] {10.3847/1538-4357/ab5e8a}, \href
  {https://ui.adsabs.harvard.edu/abs/2020ApJ...888...70K} {888, 70}

\bibitem[\protect\citeauthoryear{{Kolopanis} et~al.,}{{Kolopanis}
  et~al.}{2019}]{kolopanis_et_al_2019}
{Kolopanis} M.,  et~al., 2019, \mn@doi [\apj] {10.3847/1538-4357/ab3e3a}, \href
  {https://ui.adsabs.harvard.edu/abs/2019ApJ...883..133K} {883, 133}

\bibitem[\protect\citeauthoryear{{Lanman}, {Hazelton}, {Jacobs}, {Kolopanis},
  {Pober}, {Aguirre}  \& {Thyagarajan}}{{Lanman}
  et~al.}{2019}]{lanman_et_al_2019}
{Lanman} A.,  {Hazelton} B.,  {Jacobs} D.,  {Kolopanis} M.,  {Pober} J.,
  {Aguirre} J.,   {Thyagarajan} N.,  2019, \mn@doi [The Journal of Open Source
  Software] {10.21105/joss.01234}, \href
  {https://ui.adsabs.harvard.edu/abs/2019JOSS....4.1234L} {4, 1234}

\bibitem[\protect\citeauthoryear{{Lanman}, {Pober}, {Kern}, {de Lera Acedo},
  {DeBoer}  \& {Fagnoni}}{{Lanman} et~al.}{2020}]{lanman_et_al_2020}
{Lanman} A.~E.,  {Pober} J.~C.,  {Kern} N.~S.,  {de Lera Acedo} E.,  {DeBoer}
  D.~R.,   {Fagnoni} N.,  2020, \mn@doi [\mnras] {10.1093/mnras/staa987}, \href
  {https://ui.adsabs.harvard.edu/abs/2020MNRAS.494.3712L} {494, 3712}

\bibitem[\protect\citeauthoryear{{Li} et~al.,}{{Li}
  et~al.}{2019}]{li_et_al_2019}
{Li} W.,  et~al., 2019, \mn@doi [\apj] {10.3847/1538-4357/ab55e4}, \href
  {https://ui.adsabs.harvard.edu/abs/2019ApJ...887..141L} {887, 141}

\bibitem[\protect\citeauthoryear{{Liu} \& {Shaw}}{{Liu} \&
  {Shaw}}{2020}]{liu_and_shaw_2020}
{Liu} A.,  {Shaw} J.~R.,  2020, \mn@doi [\pasp] {10.1088/1538-3873/ab5bfd},
  \href {https://ui.adsabs.harvard.edu/abs/2020PASP..132f2001L} {132, 062001}

\bibitem[\protect\citeauthoryear{{Liu}, {Tegmark}, {Morrison}, {Lutomirski}  \&
  {Zaldarriaga}}{{Liu} et~al.}{2010}]{liu_et_al_2010}
{Liu} A.,  {Tegmark} M.,  {Morrison} S.,  {Lutomirski} A.,   {Zaldarriaga} M.,
  2010, \mn@doi [\mnras] {10.1111/j.1365-2966.2010.17174.x}, \href
  {http://adsabs.harvard.edu/abs/2010MNRAS.408.1029L} {408, 1029}

\bibitem[\protect\citeauthoryear{{Liu}, {Parsons}  \& {Trott}}{{Liu}
  et~al.}{2014}]{liu_et_al_2014a}
{Liu} A.,  {Parsons} A.~R.,   {Trott} C.~M.,  2014, \mn@doi [\prd]
  {10.1103/PhysRevD.90.023018}, \href
  {http://adsabs.harvard.edu/abs/2014PhRvD..90b3018L} {90, 023018}

\bibitem[\protect\citeauthoryear{Mardia}{Mardia}{1975}]{mardia_1975}
Mardia K.~V.,  1975, Journal of the Royal Statistical Society: Series B
  (Methodological), 37, 349

\bibitem[\protect\citeauthoryear{{Mertens} et~al.,}{{Mertens}
  et~al.}{2020}]{mertens_et_al_2020}
{Mertens} F.~G.,  et~al., 2020, \mn@doi [\mnras] {10.1093/mnras/staa327}, \href
  {https://ui.adsabs.harvard.edu/abs/2020MNRAS.493.1662M} {493, 1662}

\bibitem[\protect\citeauthoryear{{Morales}, {Beardsley}, {Pober}, {Barry},
  {Hazelton}, {Jacobs}  \& {Sullivan}}{{Morales}
  et~al.}{2019}]{morales_et_al_2019}
{Morales} M.~F.,  {Beardsley} A.,  {Pober} J.,  {Barry} N.,  {Hazelton} B.,
  {Jacobs} D.,   {Sullivan} I.,  2019, \mn@doi [\mnras]
  {10.1093/mnras/sty2844}, \href
  {https://ui.adsabs.harvard.edu/abs/2019MNRAS.483.2207M} {483, 2207}

\bibitem[\protect\citeauthoryear{{Morales}, {Pober}  \& {Hazelton}}{{Morales}
  et~al.}{2022}]{morales_et_al_2022}
{Morales} M.~F.,  {Pober} J.,   {Hazelton} B.~J.,  2022, \mn@doi [arXiv
  e-prints] {10.48550/arXiv.2211.11881}, \href
  {https://ui.adsabs.harvard.edu/abs/2022arXiv221111881M} {p. arXiv:2211.11881}

\bibitem[\protect\citeauthoryear{{Murray}}{{Murray}}{2022}]{HERAmemo113}
{Murray} S.,  2022, HERA Memo\#113

\bibitem[\protect\citeauthoryear{{Offringa}, {de Bruyn}, {Biehl}, {Zaroubi},
  {Bernardi}  \& {Pandey}}{{Offringa} et~al.}{2010}]{offringa_et_al_2010}
{Offringa} A.~R.,  {de Bruyn} A.~G.,  {Biehl} M.,  {Zaroubi} S.,  {Bernardi}
  G.,   {Pandey} V.~N.,  2010, \mn@doi [\mnras]
  {10.1111/j.1365-2966.2010.16471.x}, \href
  {http://adsabs.harvard.edu/abs/2010MNRAS.405..155O} {405, 155}

\bibitem[\protect\citeauthoryear{{Offringa}, {van de Gronde}  \&
  {Roerdink}}{{Offringa} et~al.}{2012}]{offringa_et_al_2012}
{Offringa} A.~R.,  {van de Gronde} J.~J.,   {Roerdink} J.~B.~T.~M.,  2012,
  \mn@doi [\aap] {10.1051/0004-6361/201118497}, \href
  {http://adsabs.harvard.edu/abs/2012A%26A...539A..95O} {539, A95}

\bibitem[\protect\citeauthoryear{{Offringa}, {Mertens}  \&
  {Koopmans}}{{Offringa} et~al.}{2019}]{offringa_et_al_2019}
{Offringa} A.~R.,  {Mertens} F.,   {Koopmans} L.~V.~E.,  2019, \mn@doi [\mnras]
  {10.1093/mnras/stz175}, \href
  {https://ui.adsabs.harvard.edu/abs/2019MNRAS.484.2866O} {484, 2866}

\bibitem[\protect\citeauthoryear{{Paciga} et~al.,}{{Paciga}
  et~al.}{2013}]{paciga_et_al_2013}
{Paciga} G.,  et~al., 2013, \mn@doi [\mnras] {10.1093/mnras/stt753}, \href
  {http://adsabs.harvard.edu/abs/2013MNRAS.433..639P} {433, 639}

\bibitem[\protect\citeauthoryear{{Parsons}}{{Parsons}}{2016}]{parsons_2016}
{Parsons} A.,  2016, {AIPY: Astronomical Interferometry in PYthon},
  Astrophysics Source Code Library (\mn@eprint {ascl} {1609.012})

\bibitem[\protect\citeauthoryear{{Parsons} \& {Backer}}{{Parsons} \&
  {Backer}}{2009}]{parsons_and_backer_2009}
{Parsons} A.~R.,  {Backer} D.~C.,  2009, \mn@doi [\aj]
  {10.1088/0004-6256/138/1/219}, \href
  {http://adsabs.harvard.edu/abs/2009AJ....138..219P} {138, 219}

\bibitem[\protect\citeauthoryear{{Parsons} et~al.,}{{Parsons}
  et~al.}{2010}]{parsons_et_al_2010}
{Parsons} A.~R.,  et~al., 2010, \mn@doi [\aj] {10.1088/0004-6256/139/4/1468},
  \href {http://adsabs.harvard.edu/abs/2010AJ....139.1468P} {139, 1468}

\bibitem[\protect\citeauthoryear{{Parsons}, {Pober}, {McQuinn}, {Jacobs}  \&
  {Aguirre}}{{Parsons} et~al.}{2012a}]{parsons_et_al_2012a}
{Parsons} A.,  {Pober} J.,  {McQuinn} M.,  {Jacobs} D.,   {Aguirre} J.,  2012a,
  \mn@doi [\apj] {10.1088/0004-637X/753/1/81}, \href
  {http://adsabs.harvard.edu/abs/2012ApJ...753...81P} {753, 81}

\bibitem[\protect\citeauthoryear{{Parsons}, {Pober}, {Aguirre}, {Carilli},
  {Jacobs}  \& {Moore}}{{Parsons} et~al.}{2012b}]{parsons_et_al_2012b}
{Parsons} A.~R.,  {Pober} J.~C.,  {Aguirre} J.~E.,  {Carilli} C.~L.,  {Jacobs}
  D.~C.,   {Moore} D.~F.,  2012b, \mn@doi [\apj] {10.1088/0004-637X/756/2/165},
  \href {http://adsabs.harvard.edu/abs/2012ApJ...756..165P} {756, 165}

\bibitem[\protect\citeauthoryear{{Parsons} et~al.,}{{Parsons}
  et~al.}{2014}]{parsons_et_al_2014}
{Parsons} A.~R.,  et~al., 2014, \mn@doi [\apj] {10.1088/0004-637X/788/2/106},
  \href {http://adsabs.harvard.edu/abs/2014ApJ...788..106P} {788, 106}

\bibitem[\protect\citeauthoryear{{Planck Collaboration} et~al.,}{{Planck
  Collaboration} et~al.}{2016}]{planck_2015_XIII}
{Planck Collaboration} et~al., 2016, \mn@doi [\aap]
  {10.1051/0004-6361/201525830}, \href
  {http://adsabs.harvard.edu/abs/2016A%26A...594A..13P} {594, A13}

\bibitem[\protect\citeauthoryear{{Prabu} et~al.,}{{Prabu}
  et~al.}{2015}]{prabu_et_al_2015}
{Prabu} T.,  et~al., 2015, \mn@doi [Experimental Astronomy]
  {10.1007/s10686-015-9444-3}, \href
  {https://ui.adsabs.harvard.edu/abs/2015ExA....39...73P} {39, 73}

\bibitem[\protect\citeauthoryear{{Rogers} \& {Bowman}}{{Rogers} \&
  {Bowman}}{2008}]{rogers_bowman2008}
{Rogers} A.~E.~E.,  {Bowman} J.~D.,  2008, \mn@doi [\aj]
  {10.1088/0004-6256/136/2/641}, \href
  {http://adsabs.harvard.edu/abs/2008AJ....136..641R} {136, 641}

\bibitem[\protect\citeauthoryear{{Sullivan} et~al.,}{{Sullivan}
  et~al.}{2012}]{sullivan_et_al_2012}
{Sullivan} I.~S.,  et~al., 2012, \mn@doi [\apj] {10.1088/0004-637X/759/1/17},
  \href {http://adsabs.harvard.edu/abs/2012ApJ...759...17S} {759, 17}

\bibitem[\protect\citeauthoryear{{Sutinjo}, {O'Sullivan}, {Lenc}, {Wayth},
  {Padhi}, {Hall}  \& {Tingay}}{{Sutinjo} et~al.}{2015}]{sutinjo_et_al_2015}
{Sutinjo} A.,  {O'Sullivan} J.,  {Lenc} E.,  {Wayth} R.~B.,  {Padhi} S.,
  {Hall} P.,   {Tingay} S.~J.,  2015, \mn@doi [Radio Science]
  {10.1002/2014RS005517}, \href
  {http://adsabs.harvard.edu/abs/2015RaSc...50...52S} {50, 52}

\bibitem[\protect\citeauthoryear{{Swarup}, {Ananthakrishnan}, {Kapahi}, {Rao},
  {Subrahmanya}  \& {Kulkarni}}{{Swarup} et~al.}{1991}]{swarup_et_al_1991}
{Swarup} G.,  {Ananthakrishnan} S.,  {Kapahi} V.~K.,  {Rao} A.~P.,
  {Subrahmanya} C.~R.,   {Kulkarni} V.~K.,  1991, Current Science, Vol.~60,
  NO.2/JAN25, P.~95, 1991, \href
  {http://adsabs.harvard.edu/abs/1991CuSc...60...95S} {60, 95}

\bibitem[\protect\citeauthoryear{{Tan} et~al.,}{{Tan}
  et~al.}{2021}]{tan_et_al_2021}
{Tan} J.,  et~al., 2021, \mn@doi [\apjs] {10.3847/1538-4365/ac0533}, \href
  {https://ui.adsabs.harvard.edu/abs/2021ApJS..255...26T} {255, 26}

\bibitem[\protect\citeauthoryear{{Tingay} et~al.,}{{Tingay}
  et~al.}{2013}]{tingay_et_al_2013a}
{Tingay} S.~J.,  et~al., 2013, \mn@doi [Publications of the Astronomical
  Society of Australia] {10.1017/pasa.2012.007}, \href
  {http://adsabs.harvard.edu/abs/2013PASA...30....7T} {30, 7}

\bibitem[\protect\citeauthoryear{{Trott} et~al.,}{{Trott}
  et~al.}{2020}]{trott_et_al_2020}
{Trott} C.~M.,  et~al., 2020, \mn@doi [\mnras] {10.1093/mnras/staa414}, \href
  {https://ui.adsabs.harvard.edu/abs/2020MNRAS.493.4711T} {493, 4711}

\bibitem[\protect\citeauthoryear{{Wayth} et~al.,}{{Wayth}
  et~al.}{2018}]{wayth_et_al_2018}
{Wayth} R.~B.,  et~al., 2018, \mn@doi [\pasa] {10.1017/pasa.2018.37}, \href
  {https://ui.adsabs.harvard.edu/abs/2018PASA...35...33W} {35, 33}

\bibitem[\protect\citeauthoryear{{Wilensky}, {Morales}, {Hazelton}, {Barry},
  {Byrne}  \& {Roy}}{{Wilensky} et~al.}{2019}]{wilensky_et_al_2019}
{Wilensky} M.~J.,  {Morales} M.~F.,  {Hazelton} B.~J.,  {Barry} N.,  {Byrne}
  R.,   {Roy} S.,  2019, \mn@doi [\pasp] {10.1088/1538-3873/ab3cad}, \href
  {https://ui.adsabs.harvard.edu/abs/2019PASP..131k4507W} {131, 114507}

\bibitem[\protect\citeauthoryear{{Zheng} et~al.,}{{Zheng}
  et~al.}{2014}]{zheng_et_al_2014}
{Zheng} H.,  et~al., 2014, \mn@doi [\mnras] {10.1093/mnras/stu1773}, \href
  {http://adsabs.harvard.edu/abs/2014MNRAS.445.1084Z} {445, 1084}

\bibitem[\protect\citeauthoryear{{de Oliveira-Costa}, {Tegmark}, {Gaensler},
  {Jonas}, {Landecker}  \& {Reich}}{{de Oliveira-Costa}
  et~al.}{2008}]{deOliveira-Costa_et_al_2008}
{de Oliveira-Costa} A.,  {Tegmark} M.,  {Gaensler} B.~M.,  {Jonas} J.,
  {Landecker} T.~L.,   {Reich} P.,  2008, \mn@doi [\mnras]
  {10.1111/j.1365-2966.2008.13376.x}, \href
  {http://adsabs.harvard.edu/abs/2008MNRAS.388..247D} {388, 247}

\bibitem[\protect\citeauthoryear{{van Haarlem} et~al.,}{{van Haarlem}
  et~al.}{2013}]{van_haarlem_et_al_2013}
{van Haarlem} M.~P.,  et~al., 2013, \mn@doi [\aap]
  {10.1051/0004-6361/201220873}, \href
  {http://adsabs.harvard.edu/abs/2013A%26A...556A...2V} {556, A2}

\makeatother
\end{thebibliography}




\appendix

\section{Complete Power Spectrum Values}\label{appendix:tables}
For completeness and repeatability, attached are tables of all power spectrum values, 1-$\sigma$ uncertainties derived from $P_{\text{SN}}$, and 2-$\sigma$ Upper Limits on the 21~cm power spectrum from reionization from all three baseline groups across all three spectral windows.
\begin{landscape}
    \begin{table}
    \centering
    \caption{A complete listing of all power spectrum values, 1-$\sigma$ uncertainties derived from $P_{\text{SN}}$, and upper limits on the 21cm power spectrum from the 14m baseline group.}
    \begin{tabular}{cccccccccccccc}
\multicolumn{14}{c}{14.00 m} \\ \hline \hline
\multicolumn{4}{c}{z=7.14} & & \multicolumn{4}{c}{z=6.79} & & \multicolumn{4}{c}{z=6.48} \\ 
 k $\left[\dfrac{h_{100}}{\texttt{Mpc}}\right]$ & $\Delta^{2}$ [mK$^{2}$] & $\sigma_{\Delta^{2}}$ [mK$^{2}$] &
        $\Delta_{\text{UL}}^{2}$ [mK$^{2}$] & & 
         k $\left[\dfrac{h_{100}}{\texttt{Mpc}}\right]$ & $\Delta^{2}$ [mK$^{2}$] & $\sigma_{\Delta^{2}}$ [mK$^{2}$] &
        $\Delta_{\text{UL}}^{2}$ [mK$^{2}$] & & 
         k $\left[\dfrac{h_{100}}{\texttt{Mpc}}\right]$ & $\Delta^{2}$ [mK$^{2}$] & $\sigma_{\Delta^{2}}$ [mK$^{2}$] &
        $\Delta_{\text{UL}}^{2}$ [mK$^{2}$] 
        \\
        \cline{1-4}\cline{6-9}\cline{11-14}\\
0.009 & 2.80$\times10^{5}$ & 9.47$\times10^{2}$ & 2.82$\times10^{5}$  & & 0.009 & 2.46$\times10^{5}$ & 8.07$\times10^{2}$ & 2.47$\times10^{5}$  & & 0.010 & 3.94$\times10^{5}$ & 8.44$\times10^{2}$ & 3.96$\times10^{5}$  \\ 
0.039 & 8.39$\times10^{6}$ & 5.43$\times10^{4}$ & 8.50$\times10^{6}$  & & 0.040 & 6.62$\times10^{6}$ & 4.30$\times10^{4}$ & 6.71$\times10^{6}$  & & 0.041 & 9.82$\times10^{6}$ & 3.82$\times10^{4}$ & 9.89$\times10^{6}$  \\ 
0.076 & 4.77$\times10^{6}$ & 1.24$\times10^{5}$ & 5.02$\times10^{6}$  & & 0.078 & 3.56$\times10^{6}$ & 1.00$\times10^{5}$ & 3.76$\times10^{6}$  & & 0.080 & 5.59$\times10^{6}$ & 8.16$\times10^{4}$ & 5.75$\times10^{6}$  \\ 
0.114 & 1.46$\times10^{5}$ & 4.52$\times10^{4}$ & 2.36$\times10^{5}$  & & 0.116 & 1.09$\times10^{5}$ & 3.71$\times10^{4}$ & 1.83$\times10^{5}$  & & 0.119 & 1.87$\times10^{5}$ & 2.64$\times10^{4}$ & 2.39$\times10^{5}$  \\ 
0.152 & 5.42$\times10^{3}$ & 5.17$\times10^{3}$ & 1.58$\times10^{4}$  & & 0.155 & 6.91$\times10^{3}$ & 4.17$\times10^{3}$ & 1.52$\times10^{4}$  & & 0.158 & 4.61$\times10^{3}$ & 2.69$\times10^{3}$ & 1.00$\times10^{4}$  \\ 
0.190 & 3.20$\times10^{2}$ & 2.61$\times10^{3}$ & 5.54$\times10^{3}$  & & 0.194 & 6.27$\times10^{3}$ & 2.58$\times10^{3}$ & 1.14$\times10^{4}$  & & 0.198 & 2.74$\times10^{3}$ & 1.70$\times10^{3}$ & 6.14$\times10^{3}$  \\ 
0.227 & 5.10$\times10^{3}$ & 4.50$\times10^{3}$ & 1.41$\times10^{4}$  & & 0.232 & 1.04$\times10^{4}$ & 4.02$\times10^{3}$ & 1.84$\times10^{4}$  & & 0.237 & 1.95$\times10^{3}$ & 2.94$\times10^{3}$ & 7.82$\times10^{3}$  \\ 
0.265 & 5.00$\times10^{3}$ & 7.72$\times10^{3}$ & 2.04$\times10^{4}$  & & 0.271 & 9.52$\times10^{3}$ & 6.51$\times10^{3}$ & 2.25$\times10^{4}$  & & 0.277 & 5.64$\times10^{3}$ & 4.89$\times10^{3}$ & 1.54$\times10^{4}$  \\ 
0.303 & 3.82$\times10^{4}$ & 9.22$\times10^{4}$ & 2.23$\times10^{5}$  & & 0.310 & 3.39$\times10^{4}$ & 7.57$\times10^{4}$ & 1.85$\times10^{5}$  & & 0.316 & 5.12$\times10^{4}$ & 5.40$\times10^{4}$ & 1.59$\times10^{5}$  \\ 
0.341 & 4.64$\times10^{6}$ & 1.20$\times10^{6}$ & 7.04$\times10^{6}$  & & 0.349 & 3.51$\times10^{6}$ & 9.67$\times10^{5}$ & 5.44$\times10^{6}$  & & 0.356 & 5.36$\times10^{6}$ & 7.85$\times10^{5}$ & 6.93$\times10^{6}$  \\ 
0.379 & 8.51$\times10^{7}$ & 5.44$\times10^{6}$ & 9.60$\times10^{7}$  & & 0.387 & 6.70$\times10^{7}$ & 4.29$\times10^{6}$ & 7.56$\times10^{7}$  & & 0.395 & 9.78$\times10^{7}$ & 3.79$\times10^{6}$ & 1.05$\times10^{8}$  \\ 
0.417 & 3.59$\times10^{8}$ & 1.19$\times10^{7}$ & 3.82$\times10^{8}$  & & 0.426 & 2.83$\times10^{8}$ & 9.14$\times10^{6}$ & 3.01$\times10^{8}$  & & 0.435 & 4.07$\times10^{8}$ & 8.63$\times10^{6}$ & 4.24$\times10^{8}$  \\ 
0.455 & 4.67$\times10^{8}$ & 1.54$\times10^{7}$ & 4.98$\times10^{8}$  & & 0.465 & 3.68$\times10^{8}$ & 1.19$\times10^{7}$ & 3.92$\times10^{8}$  & & 0.475 & 5.30$\times10^{8}$ & 1.12$\times10^{7}$ & 5.53$\times10^{8}$  \\ 
0.493 & 1.89$\times10^{8}$ & 1.20$\times10^{7}$ & 2.13$\times10^{8}$  & & 0.503 & 1.48$\times10^{8}$ & 9.42$\times10^{6}$ & 1.67$\times10^{8}$  & & 0.514 & 2.17$\times10^{8}$ & 8.32$\times10^{6}$ & 2.34$\times10^{8}$  \\ 
0.530 & 1.77$\times10^{7}$ & 4.51$\times10^{6}$ & 2.68$\times10^{7}$  & & 0.542 & 1.31$\times10^{7}$ & 3.64$\times10^{6}$ & 2.04$\times10^{7}$  & & 0.554 & 2.08$\times10^{7}$ & 2.95$\times10^{6}$ & 2.67$\times10^{7}$  \\ 
0.568 & 2.34$\times10^{5}$ & 6.08$\times10^{5}$ & 1.45$\times10^{6}$  & & 0.581 & 2.13$\times10^{5}$ & 5.00$\times10^{5}$ & 1.21$\times10^{6}$  & & 0.593 & 2.59$\times10^{5}$ & 3.56$\times10^{5}$ & 9.70$\times10^{5}$  \\ 
0.606 & 1.21$\times10^{5}$ & 9.17$\times10^{4}$ & 3.05$\times10^{5}$  & & 0.620 & 4.08$\times10^{4}$ & 7.39$\times10^{4}$ & 1.89$\times10^{5}$  & & 0.633 & -3.08$\times10^{4}$ & 5.82$\times10^{4}$ & 8.56$\times10^{4}$  \\ 
0.644 & 1.25$\times10^{5}$ & 1.02$\times10^{5}$ & 3.29$\times10^{5}$  & & 0.658 & -7.74$\times10^{4}$ & 8.27$\times10^{4}$ & 8.80$\times10^{4}$  & & 0.672 & 9.60$\times10^{1}$ & 6.66$\times10^{4}$ & 1.33$\times10^{5}$  \\ 
0.682 & 1.71$\times10^{5}$ & 1.21$\times10^{5}$ & 4.14$\times10^{5}$  & & 0.697 & 3.00$\times10^{3}$ & 9.84$\times10^{4}$ & 2.00$\times10^{5}$  & & 0.712 & 7.12$\times10^{4}$ & 7.91$\times10^{4}$ & 2.29$\times10^{5}$  \\ 
0.720 & 2.86$\times10^{5}$ & 1.69$\times10^{5}$ & 6.24$\times10^{5}$  & & 0.736 & 1.75$\times10^{5}$ & 1.36$\times10^{5}$ & 4.48$\times10^{5}$  & & 0.751 & 5.21$\times10^{4}$ & 1.04$\times10^{5}$ & 2.60$\times10^{5}$  \\ 
0.758 & 1.52$\times10^{6}$ & 2.27$\times10^{6}$ & 6.06$\times10^{6}$  & & 0.774 & 1.12$\times10^{6}$ & 1.87$\times10^{6}$ & 4.85$\times10^{6}$  & & 0.791 & 1.64$\times10^{6}$ & 1.33$\times10^{6}$ & 4.30$\times10^{6}$  \\ 
0.796 & 1.55$\times10^{8}$ & 2.41$\times10^{7}$ & 2.03$\times10^{8}$  & & 0.813 & 1.14$\times10^{8}$ & 1.94$\times10^{7}$ & 1.53$\times10^{8}$  & & 0.830 & 1.81$\times10^{8}$ & 1.58$\times10^{7}$ & 2.12$\times10^{8}$  \\ 
0.834 & 2.35$\times10^{9}$ & 9.17$\times10^{7}$ & 2.53$\times10^{9}$  & & 0.852 & 1.84$\times10^{9}$ & 7.22$\times10^{7}$ & 1.99$\times10^{9}$  & & 0.870 & 2.72$\times10^{9}$ & 6.39$\times10^{7}$ & 2.85$\times10^{9}$  \\ 
0.871 & 8.48$\times10^{9}$ & 1.72$\times10^{8}$ & 8.82$\times10^{9}$  & & 0.890 & 6.69$\times10^{9}$ & 1.32$\times10^{8}$ & 6.95$\times10^{9}$  & & 0.909 & 9.69$\times10^{9}$ & 1.25$\times10^{8}$ & 9.94$\times10^{9}$  \\ 
0.909 & 9.68$\times10^{9}$ & 1.96$\times10^{8}$ & 1.01$\times10^{10}$  & & 0.929 & 7.64$\times10^{9}$ & 1.50$\times10^{8}$ & 7.94$\times10^{9}$  & & 0.949 & 1.10$\times10^{10}$ & 1.42$\times10^{8}$ & 1.13$\times10^{10}$  \\ 
0.947 & 3.48$\times10^{9}$ & 1.35$\times10^{8}$ & 3.75$\times10^{9}$  & & 0.968 & 2.74$\times10^{9}$ & 1.06$\times10^{8}$ & 2.95$\times10^{9}$  & & 0.988 & 4.01$\times10^{9}$ & 9.37$\times10^{7}$ & 4.20$\times10^{9}$  \\ 
0.985 & 2.91$\times10^{8}$ & 4.58$\times10^{7}$ & 3.82$\times10^{8}$  & & & & &  & & & & & 
    \end{tabular}
    \label{tab:all_vals_14}
\end{table}
\end{landscape} 

\begin{landscape}
    \begin{table}
    \centering
        \caption{A complete listing of all power spectrum values, 1-$\sigma$ uncertainties derived from $P_{\text{SN}}$, and upper limits on the 21cm power spectrum from the 24.25~m baseline group.}
    \begin{tabular}{cccccccccccccc}
\multicolumn{14}{c}{24.25 m} \\ \hline \hline
\multicolumn{4}{c}{z=7.14} & & \multicolumn{4}{c}{z=6.79} & & \multicolumn{4}{c}{z=6.48} \\ 
 k $\left[\dfrac{h_{100}}{\texttt{Mpc}}\right]$ & $\Delta^{2}$ [mK$^{2}$] & $\sigma_{\Delta^{2}}$ [mK$^{2}$] &
        $\Delta_{\text{UL}}^{2}$ [mK$^{2}$] & & 
         k $\left[\dfrac{h_{100}}{\texttt{Mpc}}\right]$ & $\Delta^{2}$ [mK$^{2}$] & $\sigma_{\Delta^{2}}$ [mK$^{2}$] &
        $\Delta_{\text{UL}}^{2}$ [mK$^{2}$] & & 
         k $\left[\dfrac{h_{100}}{\texttt{Mpc}}\right]$ & $\Delta^{2}$ [mK$^{2}$] & $\sigma_{\Delta^{2}}$ [mK$^{2}$] &
        $\Delta_{\text{UL}}^{2}$ [mK$^{2}$] 
        \\
        \cline{1-4}\cline{6-9}\cline{11-14}\\
0.015 & 2.06$\times10^{6}$ & 4.92$\times10^{3}$ & 2.07$\times10^{6}$  & & 0.016 & 1.76$\times10^{6}$ & 4.19$\times10^{3}$ & 1.76$\times10^{6}$  & & 0.017 & 1.97$\times10^{6}$ & 4.39$\times10^{3}$ & 1.98$\times10^{6}$  \\ 
0.041 & 1.41$\times10^{7}$ & 6.24$\times10^{4}$ & 1.43$\times10^{7}$  & & 0.042 & 1.08$\times10^{7}$ & 4.98$\times10^{4}$ & 1.09$\times10^{7}$  & & 0.043 & 1.17$\times10^{7}$ & 4.47$\times10^{4}$ & 1.18$\times10^{7}$  \\ 
0.077 & 1.07$\times10^{7}$ & 1.29$\times10^{5}$ & 1.10$\times10^{7}$  & & 0.079 & 6.44$\times10^{6}$ & 1.04$\times10^{5}$ & 6.65$\times10^{6}$  & & 0.081 & 7.00$\times10^{6}$ & 8.52$\times10^{4}$ & 7.17$\times10^{6}$  \\ 
0.115 & 2.04$\times10^{6}$ & 4.59$\times10^{4}$ & 2.14$\times10^{6}$  & & 0.117 & 8.23$\times10^{5}$ & 3.78$\times10^{4}$ & 8.99$\times10^{5}$  & & 0.120 & 4.55$\times10^{5}$ & 2.69$\times10^{4}$ & 5.09$\times10^{5}$  \\ 
0.152 & 1.09$\times10^{5}$ & 5.22$\times10^{3}$ & 1.20$\times10^{5}$  & & 0.156 & 4.09$\times10^{4}$ & 4.21$\times10^{3}$ & 4.93$\times10^{4}$  & & 0.159 & 1.61$\times10^{4}$ & 2.72$\times10^{3}$ & 2.16$\times10^{4}$  \\ 
0.190 & 7.19$\times10^{3}$ & 2.63$\times10^{3}$ & 1.24$\times10^{4}$  & & 0.194 & 7.82$\times10^{3}$ & 2.59$\times10^{3}$ & 1.30$\times10^{4}$  & & 0.198 & 4.94$\times10^{3}$ & 1.71$\times10^{3}$ & 8.37$\times10^{3}$  \\ 
0.228 & 8.46$\times10^{3}$ & 4.52$\times10^{3}$ & 1.75$\times10^{4}$  & & 0.233 & 8.70$\times10^{3}$ & 4.04$\times10^{3}$ & 1.68$\times10^{4}$  & & 0.238 & 4.32$\times10^{3}$ & 2.95$\times10^{3}$ & 1.02$\times10^{4}$  \\ 
0.266 & 1.37$\times10^{4}$ & 7.74$\times10^{3}$ & 2.92$\times10^{4}$  & & 0.271 & 1.15$\times10^{4}$ & 6.53$\times10^{3}$ & 2.45$\times10^{4}$  & & 0.277 & -2.71$\times10^{3}$ & 4.91$\times10^{3}$ & 7.10$\times10^{3}$  \\ 
0.303 & 4.06$\times10^{5}$ & 9.24$\times10^{4}$ & 5.91$\times10^{5}$  & & 0.310 & 2.06$\times10^{5}$ & 7.59$\times10^{4}$ & 3.58$\times10^{5}$  & & 0.317 & 8.61$\times10^{4}$ & 5.42$\times10^{4}$ & 1.95$\times10^{5}$  \\ 
0.341 & 1.02$\times10^{7}$ & 1.20$\times10^{6}$ & 1.26$\times10^{7}$  & & 0.349 & 6.25$\times10^{6}$ & 9.69$\times10^{5}$ & 8.19$\times10^{6}$  & & 0.356 & 6.54$\times10^{6}$ & 7.87$\times10^{5}$ & 8.12$\times10^{6}$  \\ 
0.379 & 1.27$\times10^{8}$ & 5.45$\times10^{6}$ & 1.38$\times10^{8}$  & & 0.387 & 9.60$\times10^{7}$ & 4.29$\times10^{6}$ & 1.05$\times10^{8}$  & & 0.396 & 1.01$\times10^{8}$ & 3.80$\times10^{6}$ & 1.09$\times10^{8}$  \\ 
0.417 & 5.15$\times10^{8}$ & 1.19$\times10^{7}$ & 5.38$\times10^{8}$  & & 0.426 & 3.95$\times10^{8}$ & 9.15$\times10^{6}$ & 4.13$\times10^{8}$  & & 0.435 & 3.97$\times10^{8}$ & 8.64$\times10^{6}$ & 4.14$\times10^{8}$  \\ 
0.455 & 6.73$\times10^{8}$ & 1.55$\times10^{7}$ & 7.04$\times10^{8}$  & & 0.465 & 5.16$\times10^{8}$ & 1.19$\times10^{7}$ & 5.40$\times10^{8}$  & & 0.475 & 5.20$\times10^{8}$ & 1.12$\times10^{7}$ & 5.42$\times10^{8}$  \\ 
0.493 & 2.84$\times10^{8}$ & 1.20$\times10^{7}$ & 3.08$\times10^{8}$  & & 0.504 & 2.15$\times10^{8}$ & 9.42$\times10^{6}$ & 2.34$\times10^{8}$  & & 0.514 & 2.26$\times10^{8}$ & 8.33$\times10^{6}$ & 2.43$\times10^{8}$  \\ 
0.531 & 3.87$\times10^{7}$ & 4.52$\times10^{6}$ & 4.78$\times10^{7}$  & & 0.542 & 2.33$\times10^{7}$ & 3.64$\times10^{6}$ & 3.06$\times10^{7}$  & & 0.554 & 2.53$\times10^{7}$ & 2.96$\times10^{6}$ & 3.12$\times10^{7}$  \\ 
0.568 & 2.57$\times10^{6}$ & 6.08$\times10^{5}$ & 3.79$\times10^{6}$  & & 0.581 & 1.03$\times10^{6}$ & 5.00$\times10^{5}$ & 2.03$\times10^{6}$  & & 0.593 & 5.65$\times10^{5}$ & 3.56$\times10^{5}$ & 1.28$\times10^{6}$  \\ 
0.606 & 1.55$\times10^{4}$ & 9.18$\times10^{4}$ & 1.99$\times10^{5}$  & & 0.620 & 1.89$\times10^{5}$ & 7.40$\times10^{4}$ & 3.37$\times10^{5}$  & & 0.633 & 1.73$\times10^{3}$ & 5.82$\times10^{4}$ & 1.18$\times10^{5}$  \\ 
0.644 & -5.91$\times10^{4}$ & 1.02$\times10^{5}$ & 1.45$\times10^{5}$  & & 0.658 & 2.03$\times10^{5}$ & 8.28$\times10^{4}$ & 3.68$\times10^{5}$  & & 0.672 & -1.18$\times10^{4}$ & 6.66$\times10^{4}$ & 1.22$\times10^{5}$  \\ 
0.682 & -1.70$\times10^{5}$ & 1.21$\times10^{5}$ & 7.27$\times10^{4}$  & & 0.697 & 7.06$\times10^{4}$ & 9.84$\times10^{4}$ & 2.67$\times10^{5}$  & & 0.712 & -4.59$\times10^{4}$ & 7.91$\times10^{4}$ & 1.12$\times10^{5}$  \\ 
0.720 & 9.41$\times10^{4}$ & 1.69$\times10^{5}$ & 4.32$\times10^{5}$  & & 0.736 & 2.02$\times10^{5}$ & 1.36$\times10^{5}$ & 4.75$\times10^{5}$  & & 0.751 & 1.31$\times10^{5}$ & 1.04$\times10^{5}$ & 3.39$\times10^{5}$  \\ 
0.758 & 1.67$\times10^{7}$ & 2.27$\times10^{6}$ & 2.12$\times10^{7}$  & & 0.774 & 7.21$\times10^{6}$ & 1.87$\times10^{6}$ & 1.09$\times10^{7}$  & & 0.791 & 4.08$\times10^{6}$ & 1.33$\times10^{6}$ & 6.74$\times10^{6}$  \\ 
0.796 & 3.35$\times10^{8}$ & 2.41$\times10^{7}$ & 3.83$\times10^{8}$  & & 0.813 & 2.00$\times10^{8}$ & 1.95$\times10^{7}$ & 2.39$\times10^{8}$  & & 0.830 & 2.17$\times10^{8}$ & 1.58$\times10^{7}$ & 2.49$\times10^{8}$  \\ 
0.834 & 3.47$\times10^{9}$ & 9.17$\times10^{7}$ & 3.65$\times10^{9}$  & & 0.852 & 2.61$\times10^{9}$ & 7.23$\times10^{7}$ & 2.75$\times10^{9}$  & & 0.870 & 2.79$\times10^{9}$ & 6.39$\times10^{7}$ & 2.91$\times10^{9}$  \\ 
0.871 & 1.20$\times10^{10}$ & 1.72$\times10^{8}$ & 1.24$\times10^{10}$  & & 0.891 & 9.25$\times10^{9}$ & 1.32$\times10^{8}$ & 9.51$\times10^{9}$  & & 0.910 & 9.39$\times10^{9}$ & 1.25$\times10^{8}$ & 9.64$\times10^{9}$  \\ 
0.909 & 1.38$\times10^{10}$ & 1.96$\times10^{8}$ & 1.42$\times10^{10}$  & & 0.929 & 1.06$\times10^{10}$ & 1.50$\times10^{8}$ & 1.09$\times10^{10}$  & & 0.949 & 1.08$\times10^{10}$ & 1.42$\times10^{8}$ & 1.10$\times10^{10}$  \\ 
0.947 & 5.15$\times10^{9}$ & 1.35$\times10^{8}$ & 5.41$\times10^{9}$  & & 0.968 & 3.92$\times10^{9}$ & 1.06$\times10^{8}$ & 4.13$\times10^{9}$  & & 0.989 & 4.16$\times10^{9}$ & 9.37$\times10^{7}$ & 4.34$\times10^{9}$  \\ 
0.985 & 6.34$\times10^{8}$ & 4.58$\times10^{7}$ & 7.26$\times10^{8}$  & & & & &  & & & & &  
    \end{tabular}
    \label{tab:all_vals_24}
\end{table}
\end{landscape}

\begin{landscape}
    \begin{table}
    \centering
        \caption{A complete listing of all power spectrum values, 1-$\sigma$ uncertainties derived from $P_{\text{SN}}$, and upper limits on the 21cm power spectrum from the 28m baseline group.}
    \begin{tabular}{cccccccccccccc}
\multicolumn{14}{c}{28.00 m} \\ \hline \hline
\multicolumn{4}{c}{z=7.14} & & \multicolumn{4}{c}{z=6.79} & & \multicolumn{4}{c}{z=6.48} \\ 
 k $\left[\dfrac{h_{100}}{\texttt{Mpc}}\right]$ & $\Delta^{2}$ [mK$^{2}$] & $\sigma_{\Delta^{2}}$ [mK$^{2}$] &
        $\Delta_{\text{UL}}^{2}$ [mK$^{2}$] & & 
         k $\left[\dfrac{h_{100}}{\texttt{Mpc}}\right]$ & $\Delta^{2}$ [mK$^{2}$] & $\sigma_{\Delta^{2}}$ [mK$^{2}$] &
        $\Delta_{\text{UL}}^{2}$ [mK$^{2}$] & & 
         k $\left[\dfrac{h_{100}}{\texttt{Mpc}}\right]$ & $\Delta^{2}$ [mK$^{2}$] & $\sigma_{\Delta^{2}}$ [mK$^{2}$] &
        $\Delta_{\text{UL}}^{2}$ [mK$^{2}$] 
        \\
        \cline{1-4}\cline{6-9}\cline{11-14}\\
0.017 & 3.95$\times10^{6}$ & 7.57$\times10^{3}$ & 3.97$\times10^{6}$  & & 0.018 & 3.55$\times10^{6}$ & 6.45$\times10^{3}$ & 3.57$\times10^{6}$  & & 0.019 & 4.53$\times10^{6}$ & 6.75$\times10^{3}$ & 4.55$\times10^{6}$  \\ 
0.042 & 2.09$\times10^{7}$ & 6.65$\times10^{4}$ & 2.10$\times10^{7}$  & & 0.043 & 1.79$\times10^{7}$ & 5.34$\times10^{4}$ & 1.80$\times10^{7}$  & & 0.044 & 1.86$\times10^{7}$ & 4.81$\times10^{4}$ & 1.87$\times10^{7}$  \\ 
0.078 & 1.27$\times10^{7}$ & 1.31$\times10^{5}$ & 1.29$\times10^{7}$  & & 0.080 & 1.11$\times10^{7}$ & 1.06$\times10^{5}$ & 1.13$\times10^{7}$  & & 0.081 & 9.32$\times10^{6}$ & 8.69$\times10^{4}$ & 9.50$\times10^{6}$  \\ 
0.115 & 6.49$\times10^{5}$ & 4.63$\times10^{4}$ & 7.42$\times10^{5}$  & & 0.118 & 5.86$\times10^{5}$ & 3.81$\times10^{4}$ & 6.62$\times10^{5}$  & & 0.120 & 3.25$\times10^{5}$ & 2.72$\times10^{4}$ & 3.79$\times10^{5}$  \\ 
0.152 & 1.28$\times10^{4}$ & 5.25$\times10^{3}$ & 2.33$\times10^{4}$  & & 0.156 & 1.66$\times10^{4}$ & 4.23$\times10^{3}$ & 2.50$\times10^{4}$  & & 0.159 & 9.99$\times10^{3}$ & 2.74$\times10^{3}$ & 1.55$\times10^{4}$  \\ 
0.190 & 4.82$\times10^{3}$ & 2.64$\times10^{3}$ & 1.01$\times10^{4}$  & & 0.194 & 6.15$\times10^{3}$ & 2.60$\times10^{3}$ & 1.13$\times10^{4}$  & & 0.199 & 2.11$\times10^{3}$ & 1.72$\times10^{3}$ & 5.55$\times10^{3}$  \\ 
0.228 & 1.33$\times10^{4}$ & 4.53$\times10^{3}$ & 2.24$\times10^{4}$  & & 0.233 & 1.04$\times10^{4}$ & 4.05$\times10^{3}$ & 1.85$\times10^{4}$  & & 0.238 & 1.82$\times10^{3}$ & 2.96$\times10^{3}$ & 7.74$\times10^{3}$  \\ 
0.266 & 1.90$\times10^{4}$ & 7.76$\times10^{3}$ & 3.45$\times10^{4}$  & & 0.272 & 1.90$\times10^{4}$ & 6.54$\times10^{3}$ & 3.20$\times10^{4}$  & & 0.277 & 3.27$\times10^{3}$ & 4.92$\times10^{3}$ & 1.31$\times10^{4}$  \\ 
0.304 & 1.26$\times10^{5}$ & 9.25$\times10^{4}$ & 3.11$\times10^{5}$  & & 0.310 & 1.53$\times10^{5}$ & 7.60$\times10^{4}$ & 3.05$\times10^{5}$  & & 0.317 & 6.02$\times10^{4}$ & 5.43$\times10^{4}$ & 1.69$\times10^{5}$  \\ 
0.341 & 1.17$\times10^{7}$ & 1.20$\times10^{6}$ & 1.41$\times10^{7}$  & & 0.349 & 1.05$\times10^{7}$ & 9.70$\times10^{5}$ & 1.25$\times10^{7}$  & & 0.356 & 8.59$\times10^{6}$ & 7.88$\times10^{5}$ & 1.02$\times10^{7}$  \\ 
0.379 & 1.76$\times10^{8}$ & 5.45$\times10^{6}$ & 1.86$\times10^{8}$  & & 0.388 & 1.49$\times10^{8}$ & 4.30$\times10^{6}$ & 1.57$\times10^{8}$  & & 0.396 & 1.51$\times10^{8}$ & 3.80$\times10^{6}$ & 1.58$\times10^{8}$  \\ 
0.417 & 6.43$\times10^{8}$ & 1.19$\times10^{7}$ & 6.67$\times10^{8}$  & & 0.426 & 5.20$\times10^{8}$ & 9.15$\times10^{6}$ & 5.39$\times10^{8}$  & & 0.435 & 5.98$\times10^{8}$ & 8.65$\times10^{6}$ & 6.15$\times10^{8}$  \\ 
0.455 & 8.41$\times10^{8}$ & 1.55$\times10^{7}$ & 8.72$\times10^{8}$  & & 0.465 & 6.78$\times10^{8}$ & 1.19$\times10^{7}$ & 7.02$\times10^{8}$  & & 0.475 & 7.79$\times10^{8}$ & 1.12$\times10^{7}$ & 8.02$\times10^{8}$  \\ 
0.493 & 3.94$\times10^{8}$ & 1.20$\times10^{7}$ & 4.18$\times10^{8}$  & & 0.504 & 3.29$\times10^{8}$ & 9.43$\times10^{6}$ & 3.48$\times10^{8}$  & & 0.514 & 3.35$\times10^{8}$ & 8.34$\times10^{6}$ & 3.52$\times10^{8}$  \\ 
0.531 & 4.64$\times10^{7}$ & 4.52$\times10^{6}$ & 5.54$\times10^{7}$  & & 0.542 & 3.99$\times10^{7}$ & 3.64$\times10^{6}$ & 4.72$\times10^{7}$  & & 0.554 & 3.28$\times10^{7}$ & 2.96$\times10^{6}$ & 3.87$\times10^{7}$  \\ 
0.569 & 1.04$\times10^{6}$ & 6.09$\times10^{5}$ & 2.26$\times10^{6}$  & & 0.581 & 8.29$\times10^{5}$ & 5.00$\times10^{5}$ & 1.83$\times10^{6}$  & & 0.593 & 3.65$\times10^{5}$ & 3.56$\times10^{5}$ & 1.08$\times10^{6}$  \\ 
0.606 & 6.87$\times10^{4}$ & 9.18$\times10^{4}$ & 2.52$\times10^{5}$  & & 0.620 & 2.12$\times10^{4}$ & 7.40$\times10^{4}$ & 1.69$\times10^{5}$  & & 0.633 & -4.96$\times10^{3}$ & 5.83$\times10^{4}$ & 1.12$\times10^{5}$  \\ 
0.644 & -9.34$\times10^{4}$ & 1.02$\times10^{5}$ & 1.11$\times10^{5}$  & & 0.658 & 2.34$\times10^{4}$ & 8.28$\times10^{4}$ & 1.89$\times10^{5}$  & & 0.672 & -9.10$\times10^{4}$ & 6.67$\times10^{4}$ & 4.23$\times10^{4}$  \\ 
0.682 & 4.40$\times10^{4}$ & 1.21$\times10^{5}$ & 2.87$\times10^{5}$  & & 0.697 & 8.46$\times10^{4}$ & 9.85$\times10^{4}$ & 2.82$\times10^{5}$  & & 0.712 & -1.78$\times10^{5}$ & 7.91$\times10^{4}$ & -1.97$\times10^{4}$  \\ 
0.720 & 1.41$\times10^{5}$ & 1.69$\times10^{5}$ & 4.79$\times10^{5}$  & & 0.736 & 2.94$\times10^{5}$ & 1.36$\times10^{5}$ & 5.66$\times10^{5}$  & & 0.751 & -1.83$\times10^{5}$ & 1.04$\times10^{5}$ & 2.52$\times10^{4}$  \\ 
0.758 & 5.40$\times10^{6}$ & 2.27$\times10^{6}$ & 9.94$\times10^{6}$  & & 0.775 & 4.90$\times10^{6}$ & 1.87$\times10^{6}$ & 8.64$\times10^{6}$  & & 0.791 & 2.58$\times10^{6}$ & 1.33$\times10^{6}$ & 5.24$\times10^{6}$  \\ 
0.796 & 3.89$\times10^{8}$ & 2.41$\times10^{7}$ & 4.37$\times10^{8}$  & & 0.813 & 3.37$\times10^{8}$ & 1.95$\times10^{7}$ & 3.76$\times10^{8}$  & & 0.831 & 2.83$\times10^{8}$ & 1.58$\times10^{7}$ & 3.14$\times10^{8}$  \\ 
0.834 & 4.81$\times10^{9}$ & 9.17$\times10^{7}$ & 4.99$\times10^{9}$  & & 0.852 & 4.05$\times10^{9}$ & 7.23$\times10^{7}$ & 4.19$\times10^{9}$  & & 0.870 & 4.12$\times10^{9}$ & 6.39$\times10^{7}$ & 4.25$\times10^{9}$  \\ 
0.872 & 1.51$\times10^{10}$ & 1.72$\times10^{8}$ & 1.54$\times10^{10}$  & & 0.891 & 1.22$\times10^{10}$ & 1.32$\times10^{8}$ & 1.25$\times10^{10}$  & & 0.910 & 1.40$\times10^{10}$ & 1.25$\times10^{8}$ & 1.43$\times10^{10}$  \\ 
0.909 & 1.72$\times10^{10}$ & 1.96$\times10^{8}$ & 1.76$\times10^{10}$  & & 0.929 & 1.40$\times10^{10}$ & 1.50$\times10^{8}$ & 1.43$\times10^{10}$  & & 0.949 & 1.60$\times10^{10}$ & 1.42$\times10^{8}$ & 1.63$\times10^{10}$  \\ 
0.947 & 7.14$\times10^{9}$ & 1.35$\times10^{8}$ & 7.41$\times10^{9}$  & & 0.968 & 6.04$\times10^{9}$ & 1.06$\times10^{8}$ & 6.25$\times10^{9}$  & & 0.989 & 6.13$\times10^{9}$ & 9.37$\times10^{7}$ & 6.31$\times10^{9}$  \\ 
0.985 & 7.50$\times10^{8}$ & 4.58$\times10^{7}$ & 8.42$\times10^{8}$  & & & & &  & & & & &  
    \end{tabular}
    \label{tab:all_vals_27}
\end{table}
\end{landscape} 


\bsp	
\label{lastpage}
\end{document}